\documentclass[twocolumn,showpacs,aps,amsmath,amssymb,superscriptaddress,jcp]{revtex4-2}

\usepackage{adjustbox}
\usepackage{amsmath}
\usepackage{booktabs}
\usepackage{bm}
\usepackage{braket}
\usepackage{dcolumn}
\usepackage{hyperref}
\usepackage{graphicx}
\usepackage{siunitx}
\usepackage{longtable}
\usepackage[version=4]{mhchem}

\hypersetup{
    colorlinks=true,
    linkcolor=blue,
    urlcolor=black,
    citecolor=blue,
    pdfborder={0 0 0}  
}

\newcommand{\DFA}{^{\rm DFA}}
\newcommand{\GSC}{^{\rm GSC}}

\newcommand{\tot}{_{\rm tot}}
\newcommand{\atot}{_{a, \rm tot}}
\newcommand{\itot}{_{i, \rm tot}}
\newcommand{\comp}{_{\rm comp}}
\newcommand{\LUMO}{_{\rm LUMO}}
\newcommand{\HOMO}{_{\rm HOMO}}
\newcommand{\X}{_{\rm X}}
\newcommand{\XC}{_{\rm XC}}

\newcommand{\br}{\mathbf{r}}
\newcommand{\brp}{\mathbf{r^\prime}}

\newcommand{\nl}{\nonumber}

\begin{document}


\title{PyGSC: A Python tool for correcting Kohn-Sham orbital energies by mitigating the delocalization error of density functional approximations}

\author{Zipeng~An}
\affiliation{Hefei National Research Center for Physical Sciences at the Microscale \& Synergetic Innovation Center of Quantum Information and Quantum Physics, University of Science and Technology of China, Hefei, Anhui 230026, China}

\author{Xiaolong~Yang}
\affiliation{Hefei National Research Center for Physical Sciences at the Microscale \& Synergetic Innovation Center of Quantum Information and Quantum Physics, University of Science and Technology of China, Hefei, Anhui 230026, China}

\author{Xiao~Zheng} \email{xzheng@fudan.edu.cn}
\affiliation{Department of Chemistry, Fudan University, Shanghai 200438, China}

\author{Weitao~Yang} \email{weitao.yang@duke.edu}
\affiliation{Department of Chemistry, Duke University, Durham, North Carolina 27708, USA}

\date{Submitted on April~3, 2026}

\begin{abstract}
Density functional approximations (DFAs) suffer from delocalization error, which limits their accuracy in predicting electron affinities (EAs), ionization potentials (IPs), and quasiparticle energies. In this work, we present a theoretical refinement of the quasiparticle energies from density functional theory (QE-DFT) method by improving the perturbative expression for the exchange–correlation potential, leading to a more consistent description of molecular systems. We further develop an open-source Python program, PyGSC, built upon the PySCF library, which implements the modified QE-DFT framework. Benchmark tests on main-group atoms and G2/97 molecules demonstrate that the modified QE-DFT method outperforms the original DFAs, with third-order corrections achieving mean absolute deviations below $0.3$\,eV for EA and IP predictions. Application to dipole-bound states of DNA/RNA nucleobases further validates the superiority of the QE-DFT approach over original DFAs, offering an efficient and accurate approach for predicting electronic properties in large molecular systems.
\end{abstract}

\maketitle

\makeatletter
\typeout{Page width: \the\paperwidth}
\typeout{Page height: \the\paperheight}
\typeout{Text width: \the\textwidth}
\typeout{Column width: \the\columnwidth}
\typeout{Line width: \the\linewidth}
\makeatother

\section{Introduction}

Density functional theory (DFT) has emerged as one of the most widely used quantum chemical methods for exploring electronic structure, owing to its outstanding balance between accuracy and efficiency \cite{hohenberg1964inhomo, kohn1965self, parr1995density, engel2011density, jones2015density, narbe2017thirty, yang2024theoretical}. However, the exact expression for the exchange-correlation (XC) energy functional in DFT is unknown. Consequently, numerous density functional approximations (DFAs) have been proposed to evaluate this critical energy component. These approximations involve intrinsic errors, and thus inevitably result in limited accuracy for DFT in describing certain physical and chemical properties.

Enormous efforts have been devoted to mitigating errors in DFAs, employing strategies such as the incorporation of rigorous physical constraints \cite{Perdew1981, Cohen2007, sun2015strongly, hollingsworth2018can}, the design of more sophisticated functional forms \cite{perdew2001jacob, perdew2005prescription, zhang2021on, tao2003climbing, grimme2006semiempirical, schwabe2007double, goerigk2011efficient, zhang2009doubly, zhang2011a, zhang2012doubly, su2016the, Dua206902, zhang2021exploring, wang2021doubly, yan2022accurate, tan2022assessment}, and the application of machine learning techniques \cite{tozer1996exchange, zheng2004generalized, snyder2012finding, liu2017improving, nagai2018neural, nagai2020completing, wang2022improving, wang2023semilocal, chen2024development, an2025mitigating, zhang2025machine, li2025nonlocal, zheng2025machine}.

A key challenge addressed by the first approach is the systematic error in density functional calculations attributed to the systematic delocalization error of DFAs
\cite{mori2008localization, cohen2008insights, kummel2008orbital, burke2012perspective, Cohen2012challenges}. The delocalization error has a size-dependent manifestation \cite{mori2008localization,mei2022libsc}.
For small systems with small number of atoms \textit{and} with small physical extent, the delocalization error of main DFAs exhibits as violation of the Perdew–Parr–Levy–Balduz (PPLB) condition in fractional electron systems \cite{perdew1982density}. For large systems, the deviation of the total energy $E(N)$ from the piecewise linearity condition diminishes as the system size increases, ultimately vanishing in the bulk limit. However, delocalization error persists in the form of underestimated total energies for systems with integer electron numbers when an electron is added or removed. This leads to incorrect slopes in the $E(N)$ curve for large systems and at the bulk limit.

Within Kohn–Sham DFT independent of the approximations, where the XC energy is an explicit functional of the electron density, or within generalized Kohn–Sham DFT, where it depends explicitly on the first-order density matrix, these chemical potentials have been rigorously shown to correspond to the energies of the highest occupied molecular orbital (HOMO) and the lowest unoccupied molecular orbital (LUMO), respectively, which are defined as the first derivatives of the total energy with respect to electron number on either side of an integer $N$ \cite{cohen2008fractional}. 
Because of the PPLB condition, the exact chemical potentials are $-I$, the negative of ionization potential (IP) for electron removal side and $-A$, the negative of electron affinity (EA) for electron addition side. Therefore, as a direct consequence of the delocalization error, the HOMO  energy ($\epsilon_{\rm HOMO}$) underestimates $-I$ and LUMO energy ($\epsilon_{\rm LUMO}$) overestimates $-A$, leading to the underestimation of the fundamental gaps \cite{mori2008localization, perdew1983physical, zhang2000perspective, yang2000degenerate, perdew2007exchange, She2126342, Liu201036, Liu21904, Dua206902, Kraisler2025}.


To mitigate the delocalization error, Zheng et~al. developed a global scaling correction (GSC) method to systematically correct nonlinear deviations in the total energy by enforcing the PPLB condition for systems with fractional electron numbers \cite{zheng2011improving}. This approach, which initially employed a frozen orbital approximation \cite{zheng2013a} and was later extended to account for orbital relaxation effects \cite{zhang2015orbital, zheng2015scaling, zhang2018accurate}, significantly improves the prediction of molecular IP, EA, and Kohn-Sham (KS) frontier orbital energies. Implemented in a post-self-consistent-field (post-SCF) manner within the QM$^4$D software package \cite{QM4D}, the GSC method retains the high computational efficiency of the original DFA. Subsequently, Li et~al. developed the localized scaling correction (LSC) method \cite{li2015local} to address the delocalization error arising from spatially localized fractional electron distributions. This was followed by the development of the localized orbital scaling correction (LOSC) method \cite{li2017localized, su2020preserving, mei2020self}, which integrates the merits of both GSC and LSC to provide self-consistent and size-consistent corrections to the energy, energy derivatives, and electron density for systems with global or local fractional electron character. The LOSC method is implemented in an open-source software library \cite{li2017localized, su2020preserving, mei2020self}, which supports both PySCF \cite{sun2015libcint, sun2018pyscf, sun2020recent} and Psi4 \cite{smith2020psi4} platforms, ensuring both accessibility and computational efficiency for practical calculations. Following earlier work on GSC with orbital relaxation \cite{zhang2015orbital, zheng2015scaling, zhang2018accurate}, LOSC has been further developed to capture the effects of orbital relaxation, or electron screening, with a modification of the curvature of the quadratic energy correction derived based on linear response theory. The resulting linear response LOSC (lrLOSC) provides accurate description of the IP and EA for core electrons in molecules and also for band gaps in bulk systems, all within the same unified  approximation \cite{mei2021exact, yu2024accurate, williams2024correcting}.



In recent years, the GSC method has been extended to correct quasiparticle energies from DFT (QE-DFT) \cite{mei2019exc, yang2020density, kuan2024exc}, which provides a viable approach for predicting accurate excited-state energies within a ground-state framework. Despite this progress, the perturbative treatment of orbital relaxation effects in QE-DFT can encounter numerical challenges. Moreover, its implementation has so far been restricted to the QM$^4$D program. 
To address these limitations and improve the method's accuracy and accessibility, we present a revised perturbative formalism for QE-DFT in this work. We also develop a Python tool, named PyGSC, which can be integrated with the open-source PySCF package \cite{sun2015libcint, sun2018pyscf, sun2020recent}, to facilitate practical applications of the GSC method and the QE-DFT framework.

Building on these theoretical and computational advances, we systematically validate the performance of the improved approach. The benchmark datasets adopted in Refs.~\cite{zhang2015orbital, zhang2018accurate, yang2020density} are re-evaluated to quantitatively compare the prediction performance (including quasiparticle energies) between the modified and original versions of perturbative GSC methods. 

Dipole-bound states are a unique class of weakly-bound anionic states formed through the charge-dipole interactions when neutral molecules possess sufficiently large dipole moments \cite{Qian2019}. Accurately predicting these states is crucial for understanding low-energy electron-induced DNA damage, yet remains challenging for conventional DFAs due to their delocalization error, which often yields artificially unbound anions, and the strong dependence on extensive diffuse basis sets \cite{roca2008ab, tripathi2019bound}.
In this work, the modified QE-DFT method is applied to systematically study dipole-bound states of DNA/RNA nucleobases. By predicting EAs corresponding to the dipole-bound states and comparing the results with parent DFAs, we assess the capability of the QE-DFT approach in describing anionic dipole-bound states.

The remainder of this paper is organized as follows. Section~\ref{sec:method} provides a detailed exposition of the theoretical derivation of the QE-DFT method, the corrections implemented in this study, and the specifics of the program's implementation based on PySCF. Section~\ref{sec:result} presents comprehensive benchmark results, comparing the performance of the method before and after our improvements, and reports on its application and analysis in predicting the dipole-bound states of DNA and RNA nucleobases. Finally, Section~\ref{sec:conclu} offers a summary of our findings and an outlook on future directions for the development of this method.

\section{Methodology} \label{sec:method}

\subsection{Perturbative GSC approach for correcting the KS orbital energies} \label{subsec:gsc}

For an integer electron system with $N$ electrons, the exact total energy of a system with $(N+n)$ fractional electrons, $\tilde{E}\tot (N+n)$, must satisfy the PPLB condition \cite{perdew1982density, zhang2000perspective, yang2000degenerate, perdew2007exchange}.  This condition requires that the energy must be a linear interpolation between the total energies of the neighboring integer systems, $E\tot (N)$ and $E\tot(N+1)$:
\begin{equation} \label{pplb_tot}
    \tilde{E}\tot (N+n) = (1-n) E\tot (N) + n E\tot (N+1).
\end{equation}
The GSC method corrects the total energy predicted by the original DFA for the fractional electron system, $E\tot (N+n)$, by enforcing this condition. The total energy correction is given by
\begin{equation} \label{gsc_corr}
    \Delta E\GSC\tot = \tilde{E}\tot (N+n) - E\tot (N+n).
\end{equation}

Within the KS scheme, the total electronic energy is composed of four components: the KS kinetic energy, the nuclear-electron attraction energy, the Hartree energy, and the XC energy. To adhere to the PPLB condition, the GSC approach requires each individual energy component to satisfy a linear interpolation condition:
\begin{equation} \label{pplb_comp}
    \tilde{E}\comp (N+n) \equiv (1-n) E\comp (N) + nE\comp (N+1).
\end{equation}
The condition in Eq.~\eqref{pplb_comp} for component-wise correction is sufficient, but not necessary, to ensure the total energy satisfies the PPLB condition in Eq.~\eqref{pplb_tot}. In contrast, methods like LOSC \cite{li2017localized, su2020preserving, mei2020self} and GSC2 \cite{mei2021exact} aim to correct the total energy rather than each energy component individually.

The introduction of fractional electrons perturbs the electron density of the integer system. The GSC method expresses the density response as a series expansion in the fractional $n$:
\begin{align} \label{rho}
    \delta \rho (\br) & \equiv \rho^{N+n} (\br) - \rho^N (\br) \nl \\
    &= n f (\br) + n^2 \gamma (\br) + n^3 \xi (\br) + \cdots,
\end{align}
where $f (\br)$, $\gamma (\br)$, and $\xi (\br)$ are the first-, second-, and third-order Fukui functions, respectively \cite{parr1984density, yang1984electron, yang1985hardness}:
\begin{align}
    f (\br) &\equiv \lim\limits_{n \to 0} \frac{\partial \rho^{N+n} (\br)}{\partial n} \label{1f_def}, \\
    \gamma (\br) &\equiv \lim\limits_{n \to 0} \frac{1}{2} \frac{\partial^2 \rho^{N+n} (\br)}{\partial n^2} \label{2f_def}, \\
    \xi (\br) &\equiv \lim\limits_{n \to 0} \frac{1}{6} \frac{\partial^3 \rho^{N+n} (\br)}{\partial n^3}. \label{3f_def}
\end{align}
These Fukui functions are significantly affected by orbital relaxation induced by the fractional electron perturbation. The analytical expressions in terms of the density linear response functions and their computational implementation of the first-order fukui function and the chemical potentials have been developed in Ref.~\cite{mei2021exact}.
As shown in Ref.~\cite{zhang2015orbital}, the response of KS orbitals can be expressed by a similar expansion:
\begin{align} 
    \delta \phi_m (\br) &= \phi^{N+n}_m (\br) - \phi^N_m (\br) \nl \\
    &= n \delta \phi^{(1)}_m (\br) + n^2 \delta \phi^{(2)}_m (\br) + n^3 \delta \phi^{(3)}_m (\br) + \cdots. \label{eqn:or}
\end{align}
Here, $\delta \phi^{(k)}_m (\br)$ denotes the $k$th-order orbital relaxation of the $m$th KS orbital. By combining Eqs.~\eqref{1f_def}-\eqref{eqn:or}, the relations between the Fukui functions and the orbital relaxation terms are obtained: 
\begin{align}
    f (\br) &= |\phi_f (\br) |^2 + 2 \sum_{m=1}^{N_{\rm occ}} \delta \phi^{(1)}_m (\br) \phi_m (\br) \label{1f_or}, \\
    \gamma (\br) &= 2\delta \phi^{(1)}_f (\br) \phi_f (\br)  \nl \\
    &\quad + \sum_{m=1}^{N_{\rm occ}} [\delta \phi^{(1)}_m (\br) \delta \phi^{(1)}_m (\br) + 2 \delta \phi^{(2)}_m (\br) \phi_m (\br)], \label{2f_or} \\
    \xi (\br) &= [\delta \phi^{(1)}_f (\br) \delta \phi^{(1)}_f (\br) + 2 \delta \phi^{(2)}_f (\br) \phi_f (\br)] \nl \\
    &\quad  + \sum_{m=1}^{N_{\rm occ}} [\delta \phi^{(3)}_m (\br) \phi_m (\br) + \delta \phi^{(2)}_m (\br) \delta \phi^{(1)}_m (\br)]. \label{3f_or}
\end{align}
In these expressions, $\phi_f (\br)$ and $\delta \phi^{(k)}_f (\br)$ denote the frontier orbital and its $k$th-order relaxation, respectively, and $N_{\rm occ}$ denotes the number of occupied KS orbitals.


The perturbation approach for calculating Fukui functions and orbital relaxations has been detailed in Refs.~\cite{zhang2015orbital, zhang2018accurate}. In brief, the introduction of fractional electrons perturbs the KS potential, $\delta v(\br)$, which self-consistently modifies the KS orbitals and the electron density. For instance, the first-order orbital relaxation and density response are obtained through an iterative cycle:
\begin{equation} \label{scf}
    \delta \rho^{(1)}_{[0]} \to \bm{w}^{(1)}_{[0]} \to \delta \phi^{(1)}_{[0]} \to
    \delta \rho^{(1)}_{[1]} \to \bm{w}^{(1)}_{[1]} \to \delta \phi^{(1)}_{[1]} \to \cdots.
\end{equation}
Here, ${w}_{[p],mm'}^{(k)} = \bra{\phi_m} \delta v_{[p]}^{(k)} \ket{\phi_{m'} } / n^k$ represents an element of the $k$th-order perturbation Hamiltonian matrix at the $p$th iteration. Upon convergence, this cycle yields the first-order orbital relaxation and the corresponding perturbation Hamiltonian matrix. Higher-order orbital relaxation terms are subsequently obtained through analogous self-consistent cycles,  which utilize the converged Hamiltonian matrices from lower orders \cite{zheng2011improving, zheng2013a, zhang2015orbital, zheng2015scaling}.

Once the orbital relaxation terms and Fukui functions are determined, the corrected KS frontier orbital energies of the integer system can be calculated in a post-SCF manner using Janak's theorem \cite{janak1978proof}:
\begin{equation}
    \Delta \epsilon_{f}\GSC = \frac{\partial \Delta E\GSC}{\partial n} = \Delta \epsilon_{f}^{(1)} + \Delta \epsilon_{f}^{(2)} + \Delta \epsilon_{f}^{(3)} + \cdots, \label{eqn:janak}
\end{equation}
where $\Delta \epsilon_{f}^{(k)}$ is the $k$th-order correction to the KS frontier orbital energy.

This perturbative GSC formalism has been applied to the local density approximation (LDA) and generalized-gradient approximations (GGAs) such as the Perdew-Burke-Ernzerhof (PBE) \cite{perdew1996generalized} and Becke-Lee-Yang-Parr (BLYP) \cite{lee1988development, becke1988density} functionals. The approach is also applicable within the generalized KS (GKS) scheme, and the corresponding correction for the Hartree-Fock (HF) exchange energy has been derived in Refs.~\cite{zheng2013a, zhang2015orbital}. Therefore, Eq.~\eqref{eqn:janak} constitutes a general approach applicable to a wide range of DFAs, including hybrid functionals such as the B3LYP functional \cite{lee1988development, becke1988density, becke1993density} that incorporate HF exchange. Note that an equivalent approach using linear response theory lead to a first order correction to the orbital energies \cite{mei2021exact},
which is in closed analytic form and essentially the first order of Eq.~\eqref{scf}. 

\subsection{Electron affinity and ionization potential from corrected quasiparticle energies} \label{subsec:eaip}

In density functional calculations, the $\Delta$SCF method \cite{yang2024foundation, yang2024fractional, yang2024orbital, yu2025accurate} is commonly employed to compute molecular EAs and IPs:
\begin{align} \label{ai_def}
    A &= E\tot (N) - E\tot (N+1),  \nonumber \\
    I &= E\tot (N-1) - E\tot (N).
\end{align}
However, the $\Delta$SCF method requires separate SCF calculations for two different integer systems, which is computationally costly. Moreover, DFT calculations for anions often yield inaccurate results. The GSC approach provides a useful approach for the accurate calculation of vertical EAs and IPs. 

The PPLB condition, and the chemical potential theorems for ground state \cite{cohen2008fractional} established that in principle an explicit XC functional would yield the following relationships:
%
\begin{align} \label{ai_f}
    A &= - \epsilon\LUMO, \nonumber \\
    I &= - \epsilon\HOMO.
\end{align}
The GSC method effectively corrects the KS frontier orbital energies to enforce this condition. The EA and IP are thus given by:
\begin{align} \label{ai_gsc}
    A &= - (\epsilon\DFA\LUMO + \Delta \epsilon\GSC\LUMO), \nonumber \\
    I &= - (\epsilon\DFA\HOMO + \Delta \epsilon\GSC\HOMO).
\end{align}

Extending the GSC method to the QE-DFT framework \cite{mei2019quasiparticle,mei2019exc} allows the quasielectron energies $\{\omega_i\}$ and quasihole energies $\{\omega_a\}$ of an $N$-electron system to be evaluated as
\begin{align} \label{qai_def}
    \omega_a (N) \approx \epsilon_a (N) = E\atot (N+1) - E\tot (N), \nonumber \\
    \omega_i (N) \approx \epsilon_i (N) = E\tot (N) - E\itot (N-1).
\end{align}
Here, $\epsilon_a (N)$ and $\epsilon_i (N)$ denote the virtual and occupied KS orbital energies of the $N$-electron system, respectively. The quantity $E\atot (N+1)$ is the energy of the ($N+1$)-electron system formed by adding an electron to the $a$th virtual orbital, while $E\itot (N-1)$ is the energy of the ($N-1$)-electron system formed by removing an electron from the $i$th occupied orbital of the $N$-electron system. Thus, within the QE-DFT method, the vertical EAs and IPs corresponding to these quasiparticle energies are given by
\begin{align}
    A_a = -\omega_a (N) \approx -(\epsilon\DFA_a + \Delta \epsilon\GSC_a), \nl \\
    I_i = -\omega_i (N) \approx -(\epsilon\DFA_i + \Delta \epsilon\GSC_i).  \label{qai_qedft}
\end{align}

The theoretical foundation for all orbital energies as quasiparticle energies, as in Eqs.~\eqref{qai_def}-\eqref{qai_qedft},  are firmly established. For an $N$-electron system, the HOMO/LUMO orbital energies have been shown to be equal to the chemical potentials of removing/adding an electron from/to the ground states of the corresponding $(N-1)$/$(N+1)$-electron systems,and thus approximate the corresponding $-I$ and $-A$ \cite{cohen2008fractional}.
Recently it has been established that, for orbitals below HOMO and above LUMO, orbital energies for bound orbitals are the corresponding excited state chemical potentials of removing/adding an electron from/to the excited states of the corresponding $(N-1)$/$(N+1)$-electron systems and therefore approximate the corresponding excited state $-I$ and $-A$ \cite{yang2024fractional, yang2024orbital}.
This is the general chemical potential theorem for all states, within excited $\Delta$SCF theory, the theoretical foundation has been recently established \cite{yang2024foundation}.

The corrections to the virtual and occupied KS orbitals are obtained by applying the perturbative GSC formalism: 
\begin{align}
    \Delta \epsilon_{a}\GSC &= \Delta \epsilon_{a}^{(1)} + \Delta \epsilon_{a}^{(2)} + \Delta \epsilon_{a}^{(3)} + \cdots \nl \\
    \Delta \epsilon_{i}\GSC &= \Delta \epsilon_{i}^{(1)} + \Delta \epsilon_{i}^{(2)} + \Delta \epsilon_{i}^{(3)} + \cdots.
\end{align}


\subsection{Modified formula for the perturbative XC potential in the GSC method} \label{subsec:corr}

In Ref.~\cite{zhang2015orbital}, the perturbative XC potential was expressed as 
\begin{align} \label{orixc}
    \delta v\XC (\br) &= \int  K\XC (\br,\brp)  \delta \rho (\brp) \, \mathrm{d}\brp \nl \\
    &\simeq -\frac{4}{9}\, C\X [\rho(\br)]^{-\frac{2}{3}} \, \delta \rho (\br).
\end{align}
%
Here, $K\XC$ denotes the XC functional kernel, with $C\X = \frac{3}{4}(\frac{6}{\pi})^\frac{1}{3}$. Here, we explicitly consider the exchange energy of LDA, while the contribution of correlation energy is neglected because of its significantly smaller magnitude. By imposing the PPLB condition to each individual energy component, the XC potential for the $(N+n)$-electron system is a linear interpolation between neighboring integer systems:
\begin{equation} \label{corrxc}
    \tilde{v}\XC^{N+n} (\br) = (1-n) v\XC^N (\br) + n\, v\XC^{N+1} (\br).
\end{equation}

For the LDA exchange functional, the perturbative exchange potential is given by
\begin{align}
    \delta v\X (\br) &= \tilde{v}\X^{N+n} (\br) - v\X^{N} (\br) \nl \\
    &= -\frac{4}{3} n \,C\X \big\{ [\rho (\br) + f (\br) + \gamma (\br) + \xi (\br)]^{\frac{1}{3}} \nl \\
    &\qquad - [\rho (\br)]^{\frac{1}{3}} \big\}. 
\end{align}
In Ref.~\cite{zhang2015orbital}, it was presumed that at any $\br$-point in real space, the values of all orders of Fukui functions are distinctly smaller than the electron density at that point. This results in the following expressions:
\begin{align}  
    \delta v_{[p], \rm X}^{(1)} (\br) &= - \frac{4}{3} \, n C\X \big\{[\rho (\br) + f_{[p]} (\br)]^{\frac{1}{3}} - [\rho (\br)]^{\frac{1}{3}} \big\} \nl \\
    &\approx - \frac{4}{3}\, n C\X\,  [\rho (\br)]^{-\frac{2}{3}} f_{[p]} (\br), \label{eqn:deltav-1}  \\
    \delta v_{[p], \rm X}^{(2)} (\br) &= - \frac{4}{3} \, n C\X \big\{[\rho (\br) + f (\br) + \gamma_{[p]} (\br)]^{\frac{1}{3}} \nl \\
    &\qquad- [\rho (\br) + f (\br)]^{\frac{1}{3}} \big\} \nl \\
    &\approx - \frac{4}{3}\, n C\X\,  [\rho (\br)]^{-\frac{2}{3}} \gamma_{[p]} (\br).  \label{eqn:deltav-2}
\end{align}
Here, $f_{[p]} (\br)$ and $\gamma_{[p]} (\br)$ denote the first- and second-order Fukui functions at the $p$th iteration, respectively. However, in a molecular system there can be spatial regions where electron density is rather low, while the density response to the additional or removal of fractional electrons is appreciable and non-negligible. Consequently, the approximations in Eqs.~\eqref{eqn:deltav-1} and \eqref{eqn:deltav-2} no longer hold. 

We therefore propose a revised approach to evaluate the perturbative exchange potential, as detailed below. For the first-order perturbation,  the potential at the initial iteration ($p=0$) is: 
\begin{equation}
        \delta v_{[0], \rm X}^{(1)} (\br) = - \frac{4}{3} \, n C_{\rm X} \big\{[\rho (\br) + f_{[0]} (\br)]^{\frac{1}{3}} - [\rho (\br)]^{\frac{1}{3}}  \big\},  \label{eqn:deltav-3}
\end{equation}
where $f_{[0]} (\br) = |\phi_f (\br) |^2$ denotes the first-order Fukui function without orbital relaxation. For subsequent iterations ($p>0$), the potential is updated as
\begin{align}
    \delta v_{[p],\rm X}^{(1)} (\br) &= - \frac{4}{3} \, n C_{\rm X} \big\{[\rho (\br) + f_{[p]} (\br)]^{\frac{1}{3}} - [\rho (\br)]^{\frac{1}{3}}  \big\} \nl \\
    &= \delta v_{[0],\rm X}^{(1)} (\br) - \frac{4}{3} \, n C_{\rm X} \, \big\{ [\rho (\br) + f_{[p]} (\br)]^{\frac{1}{3}}  \nl \\
    &\qquad - [\rho (\br) + f_{[0]} (\br)]^{\frac{1}{3}}  \big \} \nl \\
    &\approx \delta v_{[0],\rm X}^{(1)} (\br) - \frac{4}{3} \, C_{\rm X} \, 
    \big[\rho (\br) + f_{[0]} (\br) \big]^{-\frac{2}{3}}  \nl \\
    &\qquad \times \delta \rho_{[p]}^{(1)} (\br).  \label{eqn:deltav-4}
\end{align}
Here, $\delta \rho_{[p]}^{(1)} (\br) = n[f_{[p]} (\br) - f_{[0]} (\br)]$ represents the first-order density response at the $p$th iteration. 
Similarly, the second-order perturbative potential is evaluated under the presumption that the second-order Fukui function is smaller than the first-order counterpart. For $p=0$, 
\begin{align}
    \delta v_{[0],\rm X}^{(2)} (\br) &= - \frac{4}{3} \, n C_{\rm X} \,\big\{[\rho (\br) + f (\br) + \gamma_{[0]} (\br)]^{\frac{1}{3} } \nl \\
    & \qquad - [\rho (\br) + f (\br)]^{\frac{1}{3}} \big\} \nl \\
    & \approx - \frac{4}{3} \, n C_{\rm X} \big[\rho (\br) + f (\br)\big]^{-\frac{2}{3}} \gamma_{[0]} (\br) \nl \\
    &\approx - \frac{4}{3} \, n C_{\rm X} \big[\rho (\br) + f_{[0]} (\br)\big]^{-\frac{2}{3}} 
    \gamma_{[0]} (\br),  \label{eqn:deltav-5} 
\end{align}
and for $p>0$, 
\begin{align}
    \delta v_{[p],\rm X}^{(2)} (\br) &= - \frac{4}{3} \, n C_{\rm X} \big\{[\rho (\br) + f (\br) + \gamma_{[p]} (\br)]^{\frac{1}{3}} \nl \\
    &\qquad - [\rho (\br) + f (\br)]^{\frac{1}{3}} \big\} \nl \\
    &\approx - \frac{4}{3} \, n C_{\rm X} \big[\rho (\br) + f (\br)\big]^{-\frac{2}{3}} 
    \gamma_{[p]} (\br) \nl \\
    &\approx - \frac{4}{3} \, n C_{\rm X} \big[\rho (\br) + f_{[0]} (\br)\big]^{-\frac{2}{3}} 
    \gamma_{[p]} (\br).   \label{eqn:deltav-6}
\end{align}
In these expressions, $\gamma_{[0]} (\br) = 2\delta \phi_f^{(1)} (\br) \phi_f (\br)$ denotes the second-order Fukui function without orbital relaxation. 
As demonstrated below, the refined self-consistent perturbation treatment yields improved computational accuracy for the perturbative GSC method and more consistent convergence of the perturbative series.

For GGA functionals, the GSC to the exchange energy is approximated using the LDA result, i.e., $\Delta E_{\rm X}^{\rm GSC, GGA} \approx \Delta E_{\rm X}^{\rm GSC, LDA}$. For hybrid functionals like B3LYP, the correction incorporates the HF exchange term,  $\Delta E_{\rm X}^{\rm GSC, B3LYP} \approx (1-a_0)\Delta E_{\rm X}^{\rm GSC, LDA} + a_0 \Delta E_{\rm X}^{\rm GSC, HF}$, where $a_0 = 0.20$, and $\Delta E_{\rm X}^{\rm GSC, HF}$ involves the response of the KS reduced density matrix; as detailed in Refs.~\cite{zheng2013a, zhang2015orbital}.


\subsection{An open-source Python tool for modified GSC method and QE-DFT framework} \label{subsec:code}

Complementing the theoretical refinements, this work introduces a new open-source computational tool developed to facilitate the broader adoption of the modified GSC method and QE-DFT framework in quantum chemistry. This Python-based program, designed and implemented from the ground up, retains the core QE-DFT methodology while enhancing its accessibility and performance.

The tool utilizes the PySCF package as a dependency to efficiently obtain molecular electronic structure information \cite{sun2015libcint, sun2018pyscf, sun2020recent}. PySCF provides essential capabilities including support for custom and highly diffuse basis sets, efficient computation of two-electron integrals, and robust convergence, which offer a strong foundation for our implementation.

Building on this infrastructure, our program, named PyGSC, specializes in post-SCF energy correction. It focuses on the efficient and accurate correction of KS orbital energies through third-order orbital relaxation within the QE-DFT framework. This dedicated implementation enables significant acceleration in computing quasiparticle energies while maintaining numerical accuracy comparable to high-level {\it ab initio} methods.

The PyGSC tool effectively addresses known implementation challenges and delivers improved robustness. Its post-SCF correction procedure typically yields a 2 to 9-fold speedup over conventional strategies, with the efficiency gain converging toward that of the base method as molecular size increases. PyGSC is fully developed and available as an open-source program \cite{PyGSC}.


\section{Results and Discussion} \label{sec:result}



\begin{figure*}[t]
  \centering
  \includegraphics[width=\textwidth]{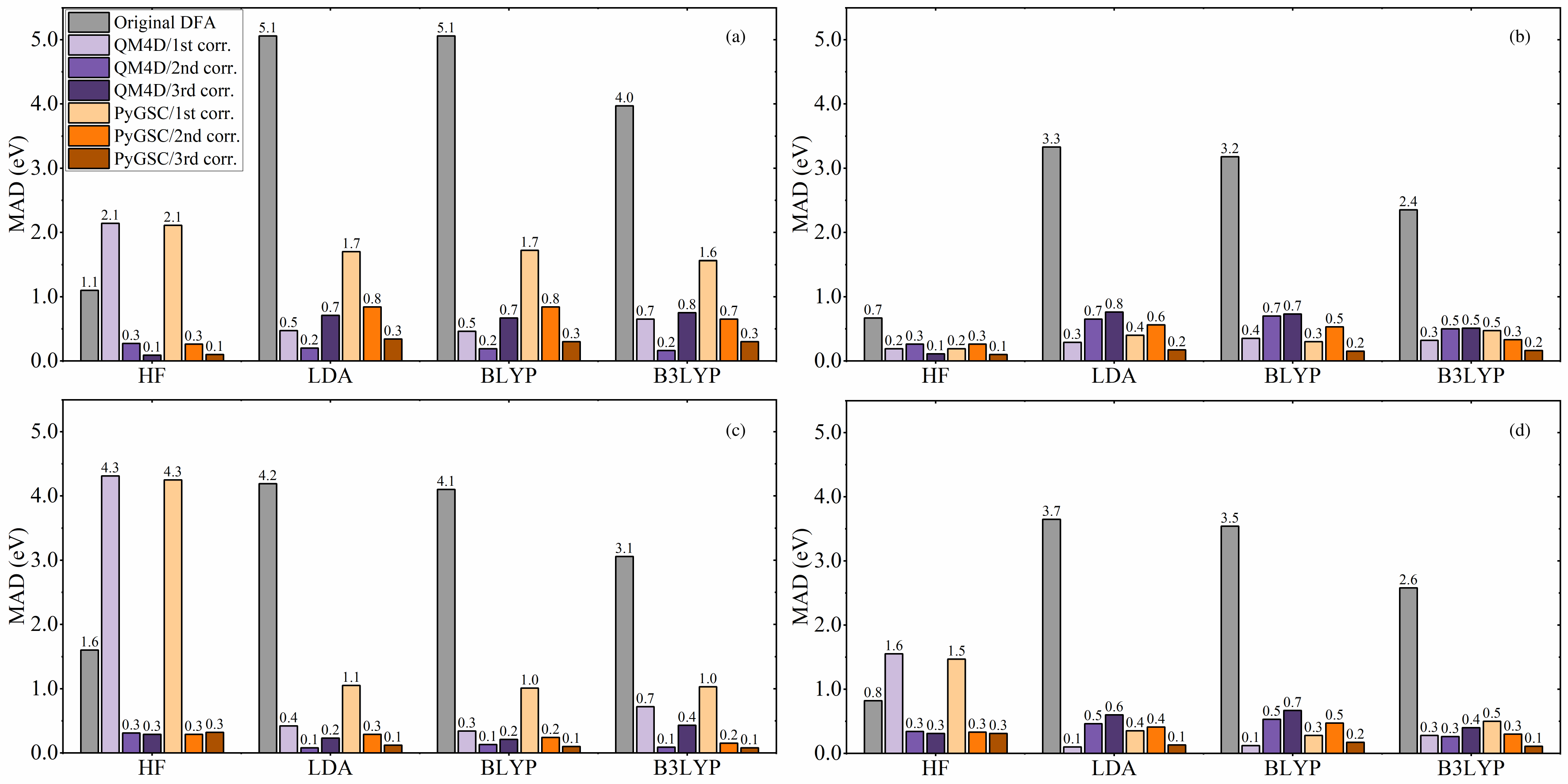} 
  \caption{ Performance comparison of various DFAs (HF, LDA, BLYP, B3LYP) for predicting atomic and molecular IPs and EAs. MADs are shown relative to $\Delta$SCF reference values. Results include:  (a) IPs for main-group atoms, (b) EAs for main-group atoms, (c) IPs for G2/97 molecules, and (d) EAs for G2/97 molecules. The main-group atomic set comprises EA calculations for 18 atoms and IP calculations for 14 atoms (excluding H, He, Ne, Ar). The G2/97 dataset contains 47 molecular EAs and 70 molecular IPs. Labels indicate correction orders: QM$^4$D/$k$th-order (original GSC method) and PyGSC/$k$th-order (modified GSC method). }
  \label{fig1}
\end{figure*}

We evaluate the performance of the modified GSC method and the PyGSC program using benchmark test sets from previous works \cite{zhang2015orbital}. Vertical EAs and IPs were calculated for the HF, LDA \cite{slater1951a, vosko1980accurate}, BLYP \cite{lee1988development, becke1988density}, and B3LYP \cite{lee1988development, becke1988density, becke1993density} methods. The test systems include EAs of 18 main-group atoms, IPs of 14 main-group atoms, and EAs and IPs for 47 and 70 molecules, respectively, from the G2/97 dataset \cite{curtiss1997assessment, curtiss1998assessment}. The 6-311++G(3df,3pd) basis set was used for all systems except the He atom, for which the 6-31G(3df,3pd) basis set was employed. The $\Delta$SCF  values from the original DFAs serve as reference. The vertical IPs and EAs predicted via Eq.~\eqref{ai_gsc} were assessed using mean absolute deviations (MADs), as shown in Fig.~\ref{fig1}.

Figure~\ref{fig1} demonstrates that the third-order corrections from the modified GSC method, implemented in PyGSC, yield consistent improvement over the original formalism of Ref.~\cite{zhang2015orbital} implemented in QM$^4$D. For the HF method, where the modifications in Section~\ref{subsec:corr} are not applicable, results from both approaches are nearly identical, with minor differences attributable to the distinct numerical convergence criteria in PySCF and QM$^4$D. In contrast, for LDA, BLYP, and B3LYP, the third-order correction given by the modified GSC method yields systematically smaller MADs. Moreover, the MADs show a consistent reduction with increasing order of orbital relaxation for most cases in Fig.~\ref{fig1}, confirming the validity of the refinements presented in Section~\ref{subsec:corr}.

We subsequently apply the modified GSC method within the QE-DFT framework to predict the EAs corresponding to the dipole-bound states of DNA/RNA nucleobases. Unlike the $\Delta$SCF approach, which faces challenges in treating these systems due to the frequent convergence to unphysical unbound states during SCF calculations for dipole-bound anions \cite{tozer2005computation, Kim2011, zhang2018accurate, yang2020density}, the QE-DFT framework enables accurate evaluation of EAs through the quasihole energies via Eq.~\eqref{qai_def}.

Given the high sensitivity of dipole-bound state energies to the diffuseness of the basis set, we employ customized, highly diffuse basis sets constructed following Jack Simons' recipe \cite{sawicka2003dipole}, starting from the aug-cc-pVTZ basis set and augmenting it with additional diffuse s, p, and d orbitals. To establish a basis set that balances accuracy and computational efficiency, we incrementally increased the number of diffuse functions until the EAs predicted by our modified QE-DFT method (including up to third-order corrections) converged to constant values. Adding highly diffuse functions to all non-hydrogen atoms can severely hinder the convergence of the SCF procedure. Therefore, in line with Ref.~\cite{tripathi2019bound}, we add extra diffuse functions only to atoms at the negative end of the molecular dipole, retaining the standard aug-cc-pVTZ basis for the remaining atoms. This strategy ensures SCF convergence stability while permitting the use of highly diffuse functions. To systematically investigate the effect of different types of diffuse orbitals, we designed ${\rm Ts}_n$, ${\rm Tsp}_n$, and ${\rm Tspd}_n$ basis sets. These are formed by adding $n$ diffuse s orbitals, $n$ diffuse s and p orbitals, and $n$ diffuse s, p, and d orbitals, respectively, to the aug-cc-pVTZ basis.

\begin{figure*}[t]
  \centering
  \includegraphics[width=0.8\textwidth]{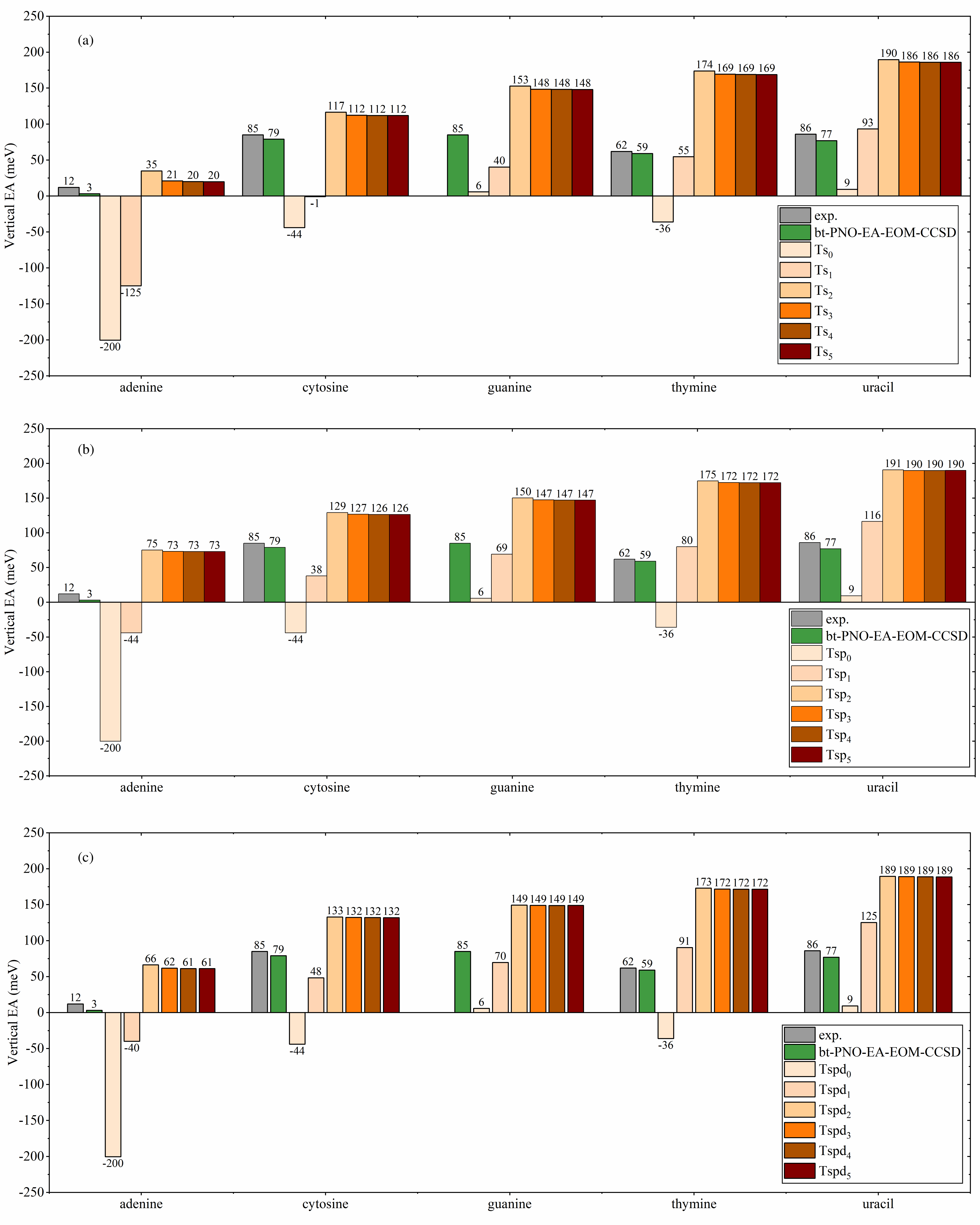} 
  \caption{EAs corresponding to the dipole-bound states of five nucleobase molecules predicted using the QE-DFT approach with (a) $\text{Ts}_n$, (b) $\text{Tsp}_n$, and (c) $\text{Tspd}_n$ basis sets, where $\text{Ts}_n = \text{aug-cc-pVTZ} + n\text{s}$, $\text{Tsp}_n = \text{aug-cc-pVTZ} + n\text{s} n\text{p}$, and $\text{Tspd}_n = \text{aug-cc-pVTZ} + n\text{s} n\text{p} n\text{d}$. Comparison includes the experimental values \cite{SCHIEDT1998511} and the results calculated by the bt-PNO-EA-EOM-CCSD method with the $\text{aug-cc-pVTZ} + 5\text{s} 5\text{p} 4\text{d}$ basis set \cite{tripathi2019bound}. Note that the experimental value for guanine is unavailable. Neutral nucleobase geometries were optimized at the B3LYP/ma-def2-TZVP level.}
  \label{fig2}
\end{figure*}

Vertical EAs corresponding to nucleobase dipole-bound states were predicted using the QE-DFT approach with the modified third-order GSC, employing B3LYP as the parent DFA via Eq.~\eqref{qai_qedft}. Figure~\ref{fig2} compares the results obtained with the three custom diffuse basis sets against experimental values \cite{SCHIEDT1998511} and the results obtained by the EA variant of equation-of-motion coupled cluster method used in singles and doubles truncation (EA-EOM-CCSD) with back-transformed pair natural orbital (bt-PNO) and the $\text{aug-cc-pVTZ} + 5\text{s} 5\text{p} 4\text{d}$ basis set \cite{tripathi2019bound}. As shown in Fig.~\ref{fig2}, the EA values calculated with the third-order corrected QE-DFT method converge to stable constants for all three basis set types as more diffuse functions are included. This convergence demonstrates the robustness of the QE-DFT framework for predicting dipole-bound anionic state energies with respect to basis set diffuseness.

\begin{figure*}[htbp]
  \centering
  \includegraphics[width=\textwidth]{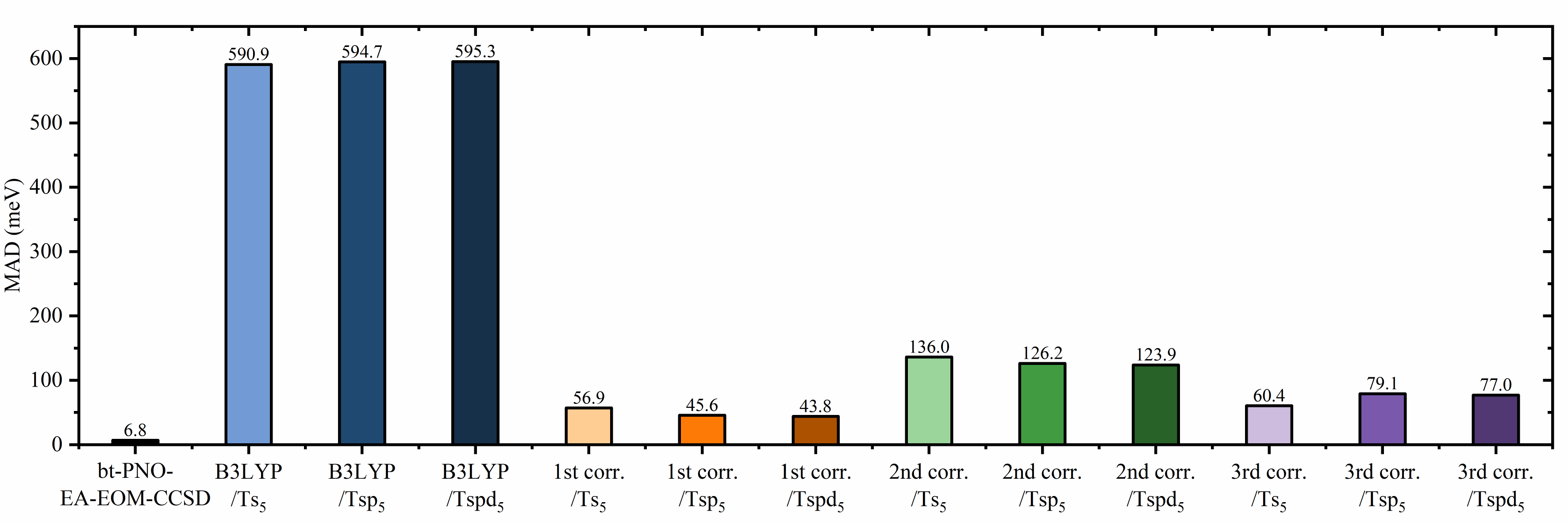} 
  \caption{MADs of EA predictions for nucleobase dipole-bound states, comparing the QE-DFT method with modified GSC at various orders against the original B3LYP functional across different basis sets. Due to unavailable experimental data for guanine, MAD statistics are based on adenine, cytosine, thymine, and uracil only. High-accuracy reference values obtained using the bt-PNO-EA-EOM-CCSD method with the aug-cc-pVTZ+5s5p4d basis set are from Ref.~\cite{tripathi2019bound}.} 
  \label{fig3}
\end{figure*}

Figure~\ref{fig3} compares the MADs of vertical EA predictions for dipole-bound states between the QE-DFT approach with the modified GSC at various orders against the original B3LYP functional, using the $\rm Ts_5$, $\rm Tsp_5$, and $\rm Tspd_5$ basis sets. The differences in MAD among these basis sets are relatively small. The first-order GSC correction within the QE-DFT framework delivers the most accurate predictions, substantially outperforming the original B3LYP. However, the MADs of the QE-DFT approach remain considerably larger than those from the high-level bt-PNO-EA-EOM-CCSD method, indicating that a parent density functional superior to B3LYP is required to provide a more accurate description of integer electron systems as reference states for correcting the energies of fractional electron systems. This also demonstrates a limitation of the GSC approach in that it does not provide any total energy correction for physical systems with integer electron numbers by its construction \cite{zheng2011improving, zheng2013a, zhang2015orbital, zheng2015scaling, zhang2018accurate}. The use of the localized orbitals in the LOSC approach potentially can overcome this limitation \cite{li2017localized, su2020preserving, mei2020self}.

Despite this limitation, the MAD values obtained with the QE-DFT method across all basis sets are significantly lower than those from the original B3LYP, demonstrating the effectiveness of the GSC correction. Although the accuracy of the QE-DFT approach does not match that of the highly accurate yet computationally expensive bt-PNO-EA-EOM-CCSD method, it offers a promising alternative for predicting EAs of dipole-bound anionic states by providing a favorable balance between computational efficiency and accuracy.

\begin{figure*}[htbp]
  \centering
  \includegraphics[width=\textwidth]{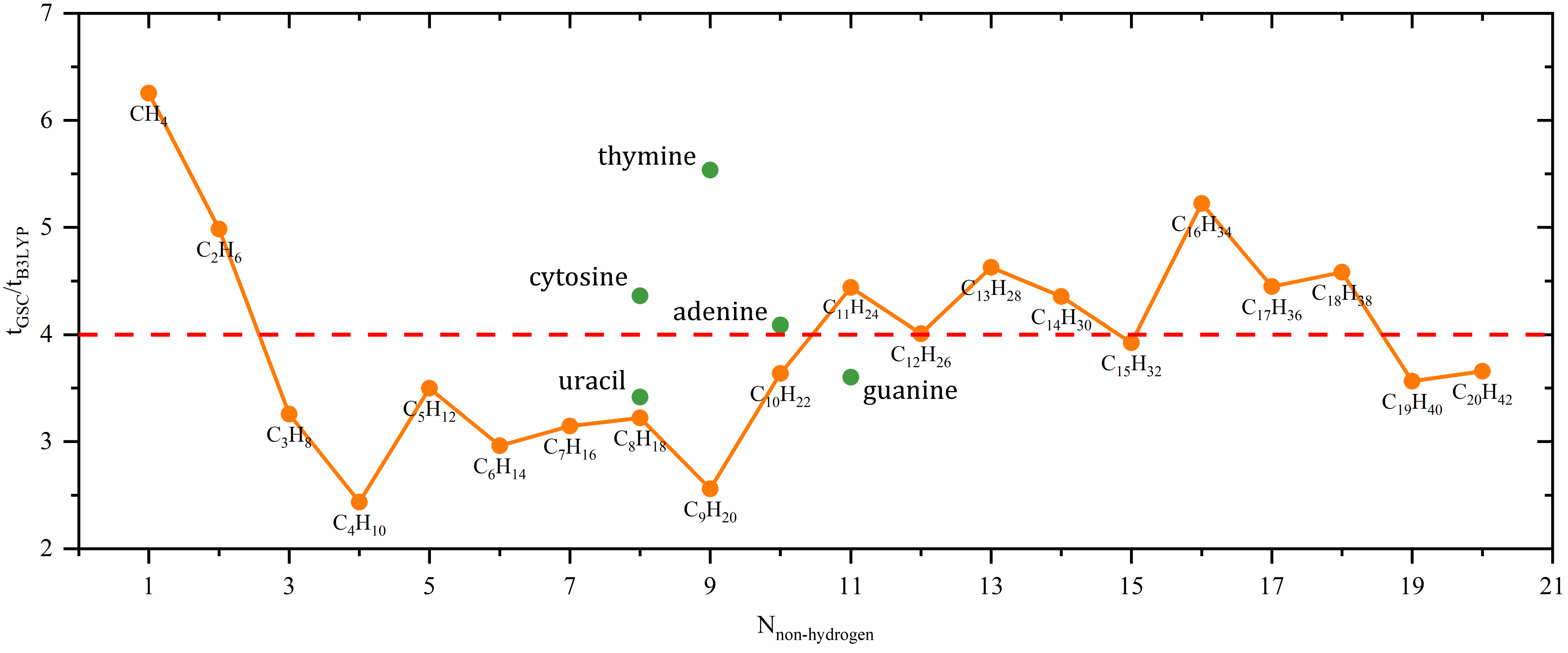} 
  \caption{Computational cost ratio $t_{\rm GSC}/t_{\rm B3LYP}$ for the post-SCF GSC calculation versus the B3LYP SCF calculation, plotted against the number of non-hydrogen atoms in n-alkanes and nucleobases.}
  \label{fig4}
\end{figure*}

Finally, we evaluate the computational efficiency of the QE-DFT method as implemented in the PyGSC program by comparing the computational time required for the post-SCF evaluation of GSC terms ($t_{\rm GSC}$) to the time for the B3LYP SCF calculation ($t_{\rm B3LYP}$). This comparison  was performed for a series of linear alkanes and DNA/RNA nucleobases. The resulting time ratios $t_{\rm GSC} / t_{\rm B3LYP}$ are summarized in Fig.~\ref{fig4}.

The results show that for linear alkane molecules, the time ratio converges to approximately 4 as the number of non-hydrogen atoms increases. A similar ratio is observed for nucleobase molecules. This demonstrates that the computational overhead introduced by PyGSC is a manageable multiple of the base B3LYP cost,  remaining acceptable compared to high-accuracy {\it ab initio} methods. The primary source of this computational overhead is the orbital relaxation step within the GSC procedure, which involves self-consistent iterations for the variable $\bm{w}^{(p)}_{[k]}$ solved via the quasi-minimal residual method \cite{freund1991qmr, freund1993a}. The stringent convergence criterion, which requires the 2-norm of $\bm{w}^{(p)}_{[k]}$ to be below $1\times 10^{-3}$ (in atomic unit) to ensure high accuracy, is the main factor contributing to the increased computation time.

\section{Concluding remarks} \label{sec:conclu}

This work presents a theoretical refinement of the perturbative GSC method and the QE-DFT framework by improving the approximate expression for the XC potential perturbation, leading to a more consistent description of total energy change and electron density response upon addition/removal of a fractional number of electrons to/from a molecular system. Alongside this theoretical development, we have implemented the modified QE-DFT method in a new open-source program, PyGSC, which is built upon the PySCF package.

The performance of the modified method and the PyGSC program was evaluated on several benchmark sets, including main-group atoms and the G2/97 dataset. The refined QE-DFT method significantly outperforms the original DFAs. It also exhibits more stable convergence behavior, with the third-order correction consistently providing superior accuracy over lower-order corrections. Furthermore, we extended the application of the QE-DFT method to predict EAs of nucleobase dipole-bound states. The results demonstrate substantial improvement over the original DFAs, offering a promising approach for predicting energies of dipole-bound anionic states in complex molecular systems. Computationally, the modified QE-DFT method incurs a manageable overhead relative to the base DFA calculation, and this relative cost remains bounded within a narrow range as the system size increases.

Despite these advances, certain limitations remain. The computational cost of the QE-DFT method, while manageable, is non-negligible, and further code optimization is possible. More importantly, the accuracy, though improved, still lags behind high-level quantum chemical methods. This may be attributed to the approximate nature of the perturbative expression for the XC potential, which is derived from the LDA and does not currently account for correlation potential perturbations. Addressing this limitation represents a key direction for future development.

\section*{Acknowledgement(s)}

Support from the National Natural Science Foundation of China (Grant Nos. 22393912, 22321003, 22425301) is acknowledged.

\section*{Data availability statement} \label{sec:data_ava}

The authors confirm that the data supporting the findings of this study are available within the main text and the Supplemental Material.

\section*{Supplemental Material} \label{sec:sm}

Supplemental Material is available online and provides additional computational details and results. This includes IPs and EAs on the benchmark tests for the parent DFAs, the original perturbative GSC method, and the modified GSC method and QE-DFT approach.


\bibliographystyle{aiptit}
\bibliography{bibrefs}

\clearpage
\onecolumngrid

\section*{Supporting Information}

\setcounter{page}{1}
\renewcommand{\thepage}{S\arabic{page}}

\setcounter{section}{0}
\setcounter{table}{0}
\setcounter{figure}{0}
\setcounter{equation}{0}

\renewcommand{\thesection}{S\arabic{section}}
\renewcommand{\thetable}{S\arabic{table}}
\renewcommand{\thefigure}{S\arabic{figure}}
\renewcommand{\theequation}{S\arabic{equation}}

\section{Computational IPs and EAs of atoms}

In this Section, calculation results of the perturbative global scaling correction (GSC) method and the quasiparticle energies from density functional theory (QE-DFT) approach for predicting ionization potentials (IPs) and electron affinities (EAs) of atoms using the HF, LDA, BLYP, and B3LYP methods are provided. All energies in tables below are in units of eV. The Gaussian atomic basis set 6-311++G(3df,3pd) is used for calculation of all atoms except that 6-31G(3df,3pd) is used for He atom. Here, $I$ and $A$ are IP and EA calculated by using the $\Delta$SCF method. We omit the contribution for degeneracies with a criterion of 0.1 eV. The mean absolute deviation (MAD) uses $I$ and $A$ as reference values.

\begin{longtable}[c]{cccccc}
  \caption{MAD of IPs for atoms with uncorrected HF and modified formalism for perturbative HF exchange potential to different orders.}\label{atm_hf_ip}\\
  \hline

  \multicolumn{1}{c}{atoms} & 
  \multicolumn{1}{c}{-I} & 
  \multicolumn{1}{c}{Original method} & 
  \multicolumn{1}{c}{1st-order correction} & 
  \multicolumn{1}{c}{2nd-order correction} & 
  \multicolumn{1}{c}{3rd-order correction} \\
  \hline
  
  \ce{H} & -13.60 & -13.60 & -13.60 & -13.60 & -13.60 \\
  \ce{He} & -23.44 & -24.87 & -20.35 & -23.13 & -23.46 \\
  \ce{Li} & -5.34 & -5.34 & -5.34 & -5.34 & -5.34 \\
  \ce{Be} & -8.04 & -8.42 & -6.81 & -7.88 & -8.14 \\
  \ce{B} & -8.03 & -8.67 & -6.74 & -7.92 & -8.09 \\
  \ce{C} & -10.79 & -11.95 & -8.53 & -10.55 & -10.90 \\
  \ce{N} & -13.89 & -15.55 & -11.06 & -13.57 & -14.01 \\
  \ce{O} & -12.03 & -14.21 & -6.76 & -11.28 & -12.39 \\
  \ce{F} & -15.67 & -18.54 & -10.30 & -14.89 & -15.98 \\
  \ce{Ne} & -19.71 & -23.20 & -14.18 & -18.90 & -19.98 \\
  \ce{Na} & -4.95 & -4.95 & -4.94 & -4.94 & -4.94 \\
  \ce{Mg} & -6.60 & -6.89 & -5.65 & -6.48 & -6.67 \\
  \ce{Al} & -5.61 & -5.94 & -4.95 & -5.56 & -5.61 \\
  \ce{Si} & -7.64 & -8.20 & -6.53 & -7.55 & -7.66 \\
  \ce{P} & -9.90 & -10.67 & -8.55 & -9.79 & -9.93 \\
  \ce{S} & -9.21 & -10.30 & -6.89 & -8.96 & -9.32 \\
  \ce{Cl} & -11.76 & -13.09 & -9.42 & -11.51 & -11.83 \\
  \ce{Ar} & -14.56 & -16.09 & -12.19 & -14.31 & -14.63 \\
  \hline\hline
  \multicolumn{1}{c}{MAD} & - & 1.10 & 2.11 & 0.26 & 0.10 \\
  \hline
\end{longtable}

\begin{longtable}[c]{cccccc}
  \caption{MAD of EAs for atoms with uncorrected HF and modified formalism for perturbative HF exchange potential to different orders.}\label{atm_hf_ea}\\
  \hline

  \multicolumn{1}{c}{atoms} & 
  \multicolumn{1}{c}{-A} & 
  \multicolumn{1}{c}{Original method} & 
  \multicolumn{1}{c}{1st-order correction} & 
  \multicolumn{1}{c}{2nd-order correction} & 
  \multicolumn{1}{c}{3rd-order correction} \\
  \hline
  
  \ce{Li} & 0.12 & 0.15 & 0.09 & 0.13 & 0.12 \\
  \ce{Be} & 0.68 & 0.80 & 0.59 & 0.72 & 0.69 \\
  \ce{B} & 0.32 & 0.80 & 0.19 & 0.55 & 0.44 \\
  \ce{C} & -0.45 & 0.50 & -0.70 & -0.05 & -0.29 \\
  \ce{N} & 1.91 & 2.40 & 1.86 & 2.13 & 2.04 \\
  \ce{O} & 0.59 & 2.00 & 0.58 & 1.26 & 0.93 \\
  \ce{F} & -1.18 & 1.08 & -1.42 & -0.29 & -0.86 \\
  \ce{Na} & 0.11 & 0.15 & 0.08 & 0.12 & 0.12 \\
  \ce{Mg} & 0.51 & 0.58 & 0.45 & 0.53 & 0.52 \\
  \ce{Al} & 0.02 & 0.38 & -0.24 & 0.14 & 0.07 \\
  \ce{Si} & -0.84 & -0.26 & -1.28 & -0.68 & -0.80 \\
  \ce{P} & 0.35 & 0.85 & 0.25 & 0.58 & 0.47 \\
  \ce{S} & -0.87 & 0.05 & -1.22 & -0.55 & -0.77 \\
  \ce{Cl} & -2.38 & -1.15 & -3.03 & -2.08 & -2.33 \\
  \hline\hline
  \multicolumn{1}{c}{MAD} & - & 0.67 & 0.19 & 0.26 & 0.10 \\
  \hline
\end{longtable}

\begin{longtable}[c]{cccccc}
  \caption{MAD of IPs for atoms with uncorrected LDA and modified formalism for perturbative LDA exchange potential to different orders.}\label{atm_lda_ip}\\
  \hline

  \multicolumn{1}{c}{atoms} & 
  \multicolumn{1}{c}{-I} & 
  \multicolumn{1}{c}{Original method} & 
  \multicolumn{1}{c}{1st-order correction} & 
  \multicolumn{1}{c}{2nd-order correction} & 
  \multicolumn{1}{c}{3rd-order correction} \\
  \hline
  
  \ce{H} & -13.02 & -7.32 & -11.45 & -12.14 & -13.37 \\
  \ce{He} & -24.25 & -15.14 & -21.52 & -22.89 & -23.65 \\
  \ce{Li} & -5.47 & -3.16 & -4.90 & -5.12 & -5.45 \\
  \ce{Be} & -9.03 & -5.60 & -7.69 & -8.17 & -8.50 \\
  \ce{B} & -8.64 & -4.12 & -6.94 & -7.65 & -8.35 \\
  \ce{C} & -11.69 & -6.14 & -9.88 & -10.85 & -11.57 \\
  \ce{N} & -15.01 & -8.43 & -13.04 & -14.25 & -14.98 \\
  \ce{O} & -14.01 & -7.47 & -11.00 & -12.41 & -13.20 \\
  \ce{F} & -17.99 & -10.40 & -15.06 & -16.72 & -17.49 \\
  \ce{Ne} & -22.24 & -13.61 & -19.31 & -21.20 & -21.98 \\
  \ce{Na} & -5.37 & -3.08 & -4.78 & -4.99 & -5.29 \\
  \ce{Mg} & -7.72 & -4.77 & -6.53 & -6.91 & -7.17 \\
  \ce{Al} & -6.03 & -3.01 & -4.99 & -5.38 & -5.74 \\
  \ce{Si} & -8.23 & -4.57 & -7.24 & -7.76 & -8.11 \\
  \ce{P} & -10.57 & -6.30 & -9.55 & -10.19 & -10.52 \\
  \ce{S} & -10.55 & -6.16 & -8.62 & -9.40 & -9.77 \\
  \ce{Cl} & -13.20 & -8.22 & -11.53 & -12.41 & -12.76 \\
  \ce{Ar} & -16.00 & -10.41 & -14.42 & -15.41 & -15.76 \\
  \hline\hline
  \multicolumn{1}{c}{MAD} & - & 5.06 & 1.70 & 0.84 & 0.34 \\
  \hline
\end{longtable}

\begin{longtable}[c]{cccccc}
  \caption{MAD of EAs for atoms with uncorrected LDA and modified formalism for perturbative LDA exchange potential to different orders.}\label{atm_lda_ea}\\
  \hline

  \multicolumn{1}{c}{atoms} & 
  \multicolumn{1}{c}{-A} & 
  \multicolumn{1}{c}{Original method} & 
  \multicolumn{1}{c}{1st-order correction} & 
  \multicolumn{1}{c}{2nd-order correction} & 
  \multicolumn{1}{c}{3rd-order correction} \\
  \hline
  
  \ce{Li} & -0.59 & -2.11 & -1.46 & -1.21 & -1.42 \\
  \ce{Be} & 0.14 & -2.09 & 0.48 & 0.84 & 0.16 \\
  \ce{B} & -0.67 & -3.93 & -0.67 & 0.01 & -0.66 \\
  \ce{C} & -1.78 & -6.06 & -2.01 & -1.08 & -1.78 \\
  \ce{N} & -0.21 & -4.39 & -0.58 & 0.47 & -0.33 \\
  \ce{O} & -1.95 & -7.17 & -2.50 & -1.17 & -1.93 \\
  \ce{F} & -4.02 & -10.28 & -4.75 & -3.18 & -3.93 \\
  \ce{Na} & -0.61 & -2.23 & -1.32 & -1.07 & -1.36 \\
  \ce{Mg} & 0.15 & -1.38 & 0.51 & 0.68 & 0.25 \\
  \ce{Al} & -0.57 & -2.91 & -0.53 & -0.14 & -0.50 \\
  \ce{Si} & -1.57 & -4.57 & -1.66 & -1.16 & -1.50 \\
  \ce{P} & -1.01 & -4.13 & -1.44 & -0.80 & -1.17 \\
  \ce{S} & -2.37 & -6.11 & -2.79 & -2.03 & -2.37 \\
  \ce{Cl} & -3.93 & -8.22 & -4.35 & -3.49 & -3.82 \\
  \hline\hline
  \multicolumn{1}{c}{MAD} & - & 3.33 & 0.40 & 0.56 & 0.17 \\
  \hline
\end{longtable}

\begin{longtable}[c]{cccccc}
  \caption{MAD of IPs for atoms with uncorrected BLYP and modified formalism for perturbative BLYP exchange potential to different orders.}\label{atm_blyp_ip}\\
  \hline

  \multicolumn{1}{c}{atoms} & 
  \multicolumn{1}{c}{-I} & 
  \multicolumn{1}{c}{Original method} & 
  \multicolumn{1}{c}{1st-order correction} & 
  \multicolumn{1}{c}{2nd-order correction} & 
  \multicolumn{1}{c}{3rd-order correction} \\
  \hline
  
  \ce{H} & -13.54 & -7.40 & -11.66 & -12.38 & -13.69 \\
  \ce{He} & -24.74 & -15.49 & -21.95 & -23.37 & -24.16 \\
  \ce{Li} & -5.53 & -3.03 & -4.80 & -5.02 & -5.38 \\
  \ce{Be} & -8.97 & -5.47 & -7.56 & -8.05 & -8.41 \\
  \ce{B} & -8.62 & -4.06 & -6.88 & -7.62 & -8.36 \\
  \ce{C} & -11.40 & -5.95 & -9.64 & -10.63 & -11.39 \\
  \ce{N} & -14.50 & -8.10 & -12.64 & -13.86 & -14.63 \\
  \ce{O} & -14.16 & -7.63 & -11.10 & -12.56 & -13.37 \\
  \ce{F} & -17.72 & -10.35 & -14.91 & -16.60 & -17.40 \\
  \ce{Ne} & -21.70 & -13.40 & -18.99 & -20.90 & -21.71 \\
  \ce{Na} & -5.36 & -2.90 & -4.58 & -4.80 & -5.12 \\
  \ce{Mg} & -7.63 & -4.57 & -6.29 & -6.68 & -6.96 \\
  \ce{Al} & -5.87 & -2.84 & -4.80 & -5.19 & -5.57 \\
  \ce{Si} & -7.94 & -4.32 & -6.93 & -7.47 & -7.84 \\
  \ce{P} & -10.18 & -5.97 & -9.19 & -9.84 & -10.19 \\
  \ce{S} & -10.40 & -6.05 & -8.48 & -9.27 & -9.65 \\
  \ce{Cl} & -12.90 & -8.03 & -11.29 & -12.18 & -12.54 \\
  \ce{Ar} & -15.60 & -10.16 & -14.11 & -15.11 & -15.47 \\
  \hline\hline
  \multicolumn{1}{c}{MAD} & - & 5.06 & 1.72 & 0.84 & 0.30 \\
  \hline
\end{longtable}

\begin{longtable}[c]{cccccc}
  \caption{MAD of EAs for atoms with uncorrected BLYP and modified formalism for perturbative BLYP exchange potential to different orders.}\label{atm_blyp_ea}\\
  \hline

  \multicolumn{1}{c}{atoms} & 
  \multicolumn{1}{c}{-A} & 
  \multicolumn{1}{c}{Original method} & 
  \multicolumn{1}{c}{1st-order correction} & 
  \multicolumn{1}{c}{2nd-order correction} & 
  \multicolumn{1}{c}{3rd-order correction} \\
  \hline
  
  \ce{Li} & -0.46 & -1.60 & -0.89 & -0.59 & -0.89 \\
  \ce{Be} & 0.30 & -1.93 & 0.60 & 0.95 & 0.17 \\
  \ce{B} & -0.35 & -3.45 & -0.27 & 0.39 & -0.36 \\
  \ce{C} & -1.31 & -5.34 & -1.39 & -0.49 & -1.25 \\
  \ce{N} & -0.25 & -4.44 & -0.68 & 0.39 & -0.44 \\
  \ce{O} & -1.70 & -6.70 & -2.11 & -0.81 & -1.62 \\
  \ce{F} & -3.55 & -9.43 & -4.02 & -2.49 & -3.29 \\
  \ce{Na} & -0.50 & -1.90 & -0.87 & -0.55 & -0.88 \\
  \ce{Mg} & 0.30 & -1.20 & 0.59 & 0.76 & 0.24 \\
  \ce{Al} & -0.28 & -2.55 & -0.24 & 0.14 & -0.24 \\
  \ce{Si} & -1.19 & -4.11 & -1.26 & -0.75 & -1.11 \\
  \ce{P} & -0.87 & -3.95 & -1.30 & -0.67 & -1.05 \\
  \ce{S} & -2.10 & -5.73 & -2.46 & -1.71 & -2.08 \\
  \ce{Cl} & -3.56 & -7.77 & -3.94 & -3.09 & -3.44 \\
  \hline\hline
  \multicolumn{1}{c}{MAD} & - & 3.18 & 0.30 & 0.53 & 0.15 \\
  \hline
\end{longtable}

\begin{longtable}[c]{cccccc}
  \caption{MAD of IPs for atoms with uncorrected B3LYP and modified formalism for perturbative B3LYP exchange potential to different orders.}\label{atm_b3lyp_ip}\\
  \hline

  \multicolumn{1}{c}{atoms} & 
  \multicolumn{1}{c}{-I} & 
  \multicolumn{1}{c}{Original method} & 
  \multicolumn{1}{c}{1st-order correction} & 
  \multicolumn{1}{c}{2nd-order correction} & 
  \multicolumn{1}{c}{3rd-order correction} \\
  \hline
  
  \ce{H} & -13.67 & -8.68 & -12.35 & -12.88 & -13.73 \\
  \ce{He} & -24.89 & -17.58 & -22.37 & -23.85 & -24.32 \\
  \ce{Li} & -5.63 & -3.57 & -5.07 & -5.24 & -5.47 \\
  \ce{Be} & -9.11 & -6.23 & -7.79 & -8.32 & -8.54 \\
  \ce{B} & -8.73 & -5.09 & -7.37 & -8.04 & -8.48 \\
  \ce{C} & -11.54 & -7.26 & -10.00 & -11.00 & -11.46 \\
  \ce{N} & -14.66 & -9.71 & -12.94 & -14.20 & -14.68 \\
  \ce{O} & -14.15 & -9.18 & -11.25 & -12.90 & -13.38 \\
  \ce{F} & -17.74 & -12.22 & -14.99 & -16.88 & -17.38 \\
  \ce{Ne} & -21.75 & -15.60 & -19.01 & -21.14 & -21.65 \\
  \ce{Na} & -5.43 & -3.40 & -4.83 & -4.99 & -5.20 \\
  \ce{Mg} & -7.72 & -5.21 & -6.51 & -6.92 & -7.10 \\
  \ce{Al} & -6.02 & -3.54 & -5.18 & -5.53 & -5.77 \\
  \ce{Si} & -8.11 & -5.20 & -7.22 & -7.75 & -7.98 \\
  \ce{P} & -10.38 & -7.04 & -9.44 & -10.10 & -10.32 \\
  \ce{S} & -10.53 & -7.11 & -8.75 & -9.63 & -9.85 \\
  \ce{Cl} & -13.06 & -9.26 & -11.49 & -12.47 & -12.69 \\
  \ce{Ar} & -15.80 & -11.59 & -14.28 & -15.37 & -15.61 \\
  \hline\hline
  \multicolumn{1}{c}{MAD} & - & 3.97 & 1.56 & 0.65 & 0.30 \\
  \hline
\end{longtable}

\begin{longtable}[c]{cccccc}
  \caption{MAD of EAs for atoms with uncorrected B3LYP and modified formalism for perturbative B3LYP exchange potential to different orders.}\label{atm_b3lyp_ea}\\
  \hline

  \multicolumn{1}{c}{atoms} & 
  \multicolumn{1}{c}{-A} & 
  \multicolumn{1}{c}{Original method} & 
  \multicolumn{1}{c}{1st-order correction} & 
  \multicolumn{1}{c}{2nd-order correction} & 
  \multicolumn{1}{c}{3rd-order correction} \\
  \hline
  
  \ce{Li} & -0.56 & -1.50 & -1.15 & -0.84 & -1.18 \\
  \ce{Be} & 0.22 & -1.43 & 0.27 & 0.61 & 0.12 \\
  \ce{B} & -0.40 & -2.69 & -0.62 & 0.06 & -0.42 \\
  \ce{C} & -1.35 & -4.29 & -1.75 & -0.82 & -1.31 \\
  \ce{N} & -0.15 & -3.18 & -0.90 & 0.13 & -0.42 \\
  \ce{O} & -1.60 & -5.16 & -2.40 & -1.09 & -1.62 \\
  \ce{F} & -3.45 & -7.59 & -4.35 & -2.77 & -3.30 \\
  \ce{Na} & -0.58 & -1.68 & -1.12 & -0.80 & -1.08 \\
  \ce{Mg} & 0.22 & -0.92 & 0.32 & 0.50 & 0.17 \\
  \ce{Al} & -0.39 & -2.10 & -0.54 & -0.12 & -0.38 \\
  \ce{Si} & -1.32 & -3.51 & -1.58 & -1.03 & -1.27 \\
  \ce{P} & -0.93 & -3.26 & -1.56 & -0.91 & -1.16 \\
  \ce{S} & -2.18 & -4.89 & -2.78 & -1.99 & -2.25 \\
  \ce{Cl} & -3.67 & -6.79 & -4.30 & -3.40 & -3.64 \\
  \hline\hline
  \multicolumn{1}{c}{MAD} & - & 2.35 & 0.47 & 0.33 & 0.16 \\
  \hline
\end{longtable}

\section{Computational IPs and EAs of G2/97 molecules}

In this Section, calculation results of unmodified and modified QE-DFT methods for predicting ionization potentials (IPs) and electron affinities (EAs) of molecules in G2-97 datasets using the HF, LDA, BLYP, and B3LYP methods are provided. All energies in tables below are in units of eV. The Gaussian atomic basis set 6-311++G(3df,3pd) is used. Here, $I$ and $A$ are IP and EA calculated by using the $\Delta$SCF method. We omit the contribution for degeneracies with a criterion of 0.1 eV. MAD uses $I$ and $A$ as reference values.

Notably that for IPs of $2A_1$ cation of \ce{H2S} or $2\Pi_u$ cation of \ce{N2}, the corresponding $I$ is compared to $\epsilon_{\rm HOMO-1}$ or $\epsilon_{\rm HOMO-2}$.

\begin{longtable}[c]{cccccc}
  \caption{MAD of IPs for molecules in G2-97 datasets with uncorrected HF and modified formalism for perturbative HF exchange potential to different orders. The \ce{HOF} molecule has been excluded.}\label{g2_hf_ip}\\
  \hline

  \multicolumn{1}{c}{molecules} & 
  \multicolumn{1}{c}{-I} & 
  \multicolumn{1}{c}{Original method} & 
  \multicolumn{1}{c}{1st-order correction} & 
  \multicolumn{1}{c}{2nd-order correction} & 
  \multicolumn{1}{c}{3rd-order correction} \\
  \hline
  
  \ce{H4C} & -13.23 & -14.80 & -10.90 & -13.03 & -13.28 \\
  \ce{H3N} & -9.46 & -11.77 & -3.43 & -8.60 & -9.84 \\
  \ce{HO} & -11.35 & -13.97 & -5.26 & -10.44 & -11.71 \\
  \ce{H2O} & -10.96 & -13.88 & -4.22 & -9.92 & -11.33 \\
  \ce{HF} & -14.28 & -17.68 & -7.91 & -13.28 & -14.56 \\
  \ce{H4Si} & -11.93 & -13.21 & -9.71 & -11.91 & -12.01 \\
  \ce{HP} & -9.65 & -10.51 & -8.23 & -9.54 & -9.67 \\
  \ce{H2P} & -9.32 & -10.25 & -7.82 & -9.21 & -9.33 \\
  \ce{H3P} & -9.43 & -10.58 & -7.30 & -9.24 & -9.49 \\
  \ce{HS} & -9.15 & -10.35 & -6.70 & -8.89 & -9.25 \\
  \ce{H2S} (\ce{^2B_1}) & -9.19 & -10.48 & -6.67 & -8.92 & -9.29 \\
  \ce{H2S} (\ce{^2A_1}) & -12.32 & -13.61 & -10.13 & -12.14 & -12.36 \\
  \ce{HCl} & -11.51 & -12.98 & -9.00 & -11.23 & -11.58 \\
  \ce{H2C2} & -9.78 & -11.18 & -7.55 & -9.51 & -9.85 \\
  \ce{H4C2} & -8.93 & -10.29 & -6.83 & -8.69 & -8.98 \\
  \ce{CO} & -13.09 & -15.14 & -9.65 & -13.05 & -13.30 \\
  \ce{N2} (\ce{^2\Sigma_g}) & -15.17 & -16.64 & -13.13 & -14.96 & -15.22 \\
  \ce{N2} (\ce{^2\Pi_u}) & -15.64 & -17.27 & -13.16 & -15.45 & -15.70 \\
  \ce{O2} & -13.50 & -15.34 & -10.92 & -13.26 & -13.57 \\
  \ce{P2} & -9.17 & -10.08 & -7.74 & -9.04 & -9.21 \\
  \ce{S2} & -9.66 & -10.52 & -8.42 & -9.60 & -9.68 \\
  \ce{Cl2} & -11.25 & -12.25 & -9.81 & -11.18 & -11.28 \\
  \ce{FCl} & -12.12 & -13.57 & -10.10 & -12.04 & -12.16 \\
  \ce{CS} & -9.89 & -12.55 & 2.02 & -10.98 & -13.08 \\
  \ce{BF3} & -15.24 & -18.09 & -14.24 & -16.40 & -16.52 \\
  \ce{BCl3} & -11.24 & -12.51 & -10.69 & -11.72 & -11.75 \\
  \ce{B2F4} & -13.05 & -14.36 & -11.39 & -13.04 & -13.06 \\
  \ce{CO2} & -12.36 & -14.83 & -5.96 & -12.14 & -12.36 \\
  \ce{CF2} & -11.53 & -13.19 & -8.35 & -11.27 & -11.64 \\
  \ce{COS} & -10.08 & -11.46 & -6.75 & -9.94 & -10.23 \\
  \ce{CS2} & -8.70 & -10.16 & -2.64 & -8.54 & -8.53 \\
  \ce{H2C} & -9.70 & -11.14 & -7.13 & -9.43 & -9.76 \\
  \ce{H3C} & -8.97 & -10.46 & -6.22 & -8.67 & -9.02 \\
  \ce{H5C2} & -7.94 & -9.52 & -5.65 & -7.84 & -7.90 \\
  \ce{H4C3} (cyc) & -8.49 & -9.73 & -6.92 & -8.41 & -8.50 \\
  \ce{H4C3} (all) & -8.70 & -10.31 & 41.44 & -8.20 & -14.46 \\
  \ce{H7C3} & -7.20 & -8.89 & -5.06 & -7.19 & -7.14 \\
  \ce{H6C6} & -7.96 & -9.16 & -5.44 & -7.79 & -7.98 \\
  \ce{H8C7} & -7.53 & -8.81 & -4.89 & -7.49 & -6.67 \\
  \ce{CN} & -12.47 & -14.19 & -9.71 & -12.46 & -13.01 \\
  \ce{HCO} & -9.66 & -10.80 & -7.86 & -9.53 & -9.65 \\
  \ce{CH2OH} & -7.95 & -9.41 & -5.72 & -7.79 & -7.99 \\
  \ce{CH3O} & -9.75 & -12.18 & -6.69 & -9.89 & -9.53 \\
  \ce{H4CO} & -9.70 & -12.29 & -6.68 & -9.92 & -9.36 \\
  \ce{H3CF} & -12.39 & -14.43 & -11.07 & -12.93 & -12.97 \\
  \ce{H2CS} & -8.04 & -9.60 & -5.07 & -8.00 & -8.21 \\
  \ce{CH2SH} & -7.58 & -8.60 & -5.84 & -7.44 & -7.56 \\
  \ce{H4CS} & -8.30 & -9.75 & -5.87 & -8.11 & -8.33 \\
  \ce{H3CCl} & -10.25 & -11.88 & -8.02 & -10.16 & -10.21 \\
  \ce{H6C2O} & -9.48 & -12.04 & -7.62 & -10.13 & -9.11 \\
  \ce{H4C2O} & -8.84 & -11.64 & -1.74 & -8.95 & -8.45 \\
  \ce{H3COF} & -10.41 & -13.49 & 17.69 & -11.51 & -9.44 \\
  \ce{H4C2S} & -7.82 & -9.44 & -5.22 & -7.62 & -7.84 \\
  \ce{C2N2} & -11.91 & -13.48 & 3.72 & -11.35 & -12.19 \\
  \ce{H4C4O} & -7.52 & -8.73 & -3.62 & -7.27 & -7.65 \\
  \ce{H5C4N} & -6.87 & -8.12 & -3.61 & -6.64 & -6.97 \\
  \ce{H6C6O} & -7.34 & -8.62 & -2.21 & -7.50 & -8.06 \\
  \ce{H7C6N} & -6.69 & -8.04 & 2.13 & -7.13 & -8.16 \\
  \ce{H4B2} & -8.84 & -10.06 & -6.79 & -8.64 & -8.89 \\
  \ce{HN} & -12.84 & -14.89 & -9.39 & -12.42 & -12.95 \\
  \ce{H2N} & -10.51 & -12.66 & -5.23 & -9.81 & -10.80 \\
  \ce{H2N2} & -9.76 & -11.22 & -7.19 & -9.53 & -9.85 \\
  \ce{H3N2} & -8.49 & -10.04 & -5.00 & -7.97 & -8.23 \\
  \ce{HOF} & -11.74 & -14.92 & 3019.45 & -11.03 & 301.71 \\
  \ce{H2Si} & -8.41 & -9.25 & -6.51 & -8.24 & -8.47 \\
  \ce{H3Si} & -8.46 & -9.22 & -7.41 & -8.43 & -8.47 \\
  \ce{H2Si2} & -7.09 & -7.98 & -4.97 & -6.98 & -7.11 \\
  \ce{H4Si2} & -6.91 & -7.71 & -5.02 & -6.74 & -6.96 \\
  \ce{H5Si2} & -7.94 & -8.88 & -6.64 & -7.91 & -7.90 \\
  \ce{H6Si2} & -9.82 & -11.05 & -8.18 & -9.77 & -9.82 \\
  \hline\hline
  \multicolumn{1}{c}{MAD} & - & 1.60 & 4.25 & 0.29 & 0.32 \\
  \hline
\end{longtable}

\begin{longtable}[c]{cccccc}
  \caption{MAD of EAs for molecules in G2-97 datasets with uncorrected HF and modified formalism for perturbative HF exchange potential to different orders.}\label{g2_hf_ea}\\
  \hline

  \multicolumn{1}{c}{molecules} & 
  \multicolumn{1}{c}{-A} & 
  \multicolumn{1}{c}{Original method} & 
  \multicolumn{1}{c}{1st-order correction} & 
  \multicolumn{1}{c}{2nd-order correction} & 
  \multicolumn{1}{c}{3rd-order correction} \\
  \hline
  
  \ce{HC} & -0.35 & 0.56 & -0.61 & -0.02 & -0.26 \\
  \ce{H2C} & 1.19 & 1.36 & 1.34 & 1.35 & 1.35 \\
  \ce{H3C} & 1.16 & 1.29 & 1.28 & 1.28 & 1.28 \\
  \ce{HN} & 1.50 & 1.31 & 1.22 & 1.27 & 1.27 \\
  \ce{H2N} & 1.09 & 1.20 & 1.14 & 1.18 & 1.17 \\
  \ce{HO} & 0.30 & 1.24 & 1.16 & 1.21 & 1.20 \\
  \ce{HSi} & -0.75 & -0.18 & -1.16 & -0.60 & -0.71 \\
  \ce{H2Si} & -0.58 & -0.05 & -0.95 & -0.45 & -0.55 \\
  \ce{H3Si} & 0.25 & 0.72 & 0.15 & 0.46 & 0.36 \\
  \ce{HP} & 0.18 & 0.74 & 0.05 & 0.43 & 0.30 \\
  \ce{H2P} & -0.05 & 0.59 & -0.21 & 0.22 & 0.07 \\
  \ce{HS} & -1.07 & -0.07 & -1.44 & -0.73 & -0.97 \\
  \ce{O2} & 0.96 & 2.44 & 0.26 & 1.35 & 1.05 \\
  \ce{NO} & 0.58 & 1.99 & -13.65 & 0.58 & 1.60 \\
  \ce{CN} & -2.91 & -0.92 & -3.94 & -2.22 & -2.77 \\
  \ce{OP} & -0.84 & -0.09 & -3.10 & -0.71 & -0.53 \\
  \ce{S2} & -0.81 & 0.02 & -1.61 & -0.68 & -0.80 \\
  \ce{Cl2} & -0.59 & 0.33 & -1.20 & -0.39 & -0.61 \\
  \ce{C2} & -4.56 & -3.17 & -5.81 & -4.38 & -4.52 \\
  \ce{C2O} & -0.85 & 0.15 & -1.73 & -0.64 & -0.82 \\
  \ce{CF2} & 0.64 & 1.69 & 0.62 & 1.18 & 0.84 \\
  \ce{CNO} & -2.09 & -0.60 & -5.15 & -1.91 & -2.01 \\
  \ce{NO2} & -0.96 & 0.33 & -1.64 & -0.59 & -0.93 \\
  \ce{O3} & -2.90 & -1.21 & -24.87 & -3.61 & -0.17 \\
  \ce{OF} & -0.52 & 1.60 & -1.63 & 0.18 & -0.40 \\
  \ce{O2S} & -1.15 & -0.12 & -2.20 & -1.04 & -1.17 \\
  \ce{OS2} & -2.05 & -1.04 & -3.84 & -1.86 & -1.78 \\
  \ce{HC2} & -1.54 & -0.52 & -3.62 & -1.45 & -1.78 \\
  \ce{H3C2} & 1.30 & 1.26 & 1.24 & 1.25 & 1.25 \\
  \ce{H2C3} & -1.26 & -0.04 & -3.19 & -0.97 & -1.13 \\
  \ce{H3C3} & 0.81 & 1.07 & 1.02 & 1.05 & 1.04 \\
  \ce{H5C3} & 1.26 & 1.17 & 1.16 & 1.16 & 1.16 \\
  \ce{HCO} & 1.10 & 1.21 & 1.04 & 1.14 & 1.14 \\
  \ce{HCF} & 0.31 & 1.25 & 0.24 & 0.75 & 0.51 \\
  \ce{CH3O} & 0.47 & 1.09 & 1.05 & 1.07 & 1.07 \\
  \ce{H3CS} & 1.10 & 1.20 & 1.14 & 1.17 & 1.17 \\
  \ce{H2CS} & 0.04 & 0.96 & -2.14 & 0.42 & 0.59 \\
  \ce{CH2CN} & -0.08 & 1.01 & 0.89 & 0.96 & 0.96 \\
  \ce{CH2NC} & 0.67 & 1.06 & 0.92 & 1.01 & 1.00 \\
  \ce{HC2O} & -0.96 & 0.23 & -3.12 & -0.74 & -0.89 \\
  \ce{CH2CHO} & -0.08 & 1.16 & 1.01 & 1.11 & 1.10 \\
  \ce{CH3CO} & 1.20 & 0.86 & 0.78 & 0.83 & 0.82 \\
  \ce{H5C2O} & 0.24 & 1.04 & 0.98 & 1.01 & 1.01 \\
  \ce{H5C2S} & -0.69 & 0.58 & -1.10 & -0.24 & -0.59 \\
  \ce{HLi} & -0.25 & -0.21 & -0.31 & -0.25 & -0.25 \\
  \ce{HNO} & 0.35 & 1.18 & 1.01 & 1.11 & 1.11 \\
  \ce{HO2} & 0.82 & 1.21 & 1.05 & 1.14 & 1.13 \\
  \hline\hline
  \multicolumn{1}{c}{MAD} & - & 0.82 & 1.47 & 0.33 & 0.31 \\
  \hline
\end{longtable}

\begin{longtable}[c]{cccccc}
  \caption{MAD of IPs for molecules in G2-97 datasets with uncorrected LDA and modified formalism for perturbative LDA exchange potential to different orders.}\label{g2_lda_ip}\\
  \hline

  \multicolumn{1}{c}{molecules} & 
  \multicolumn{1}{c}{-I} & 
  \multicolumn{1}{c}{Original method} & 
  \multicolumn{1}{c}{1st-order correction} & 
  \multicolumn{1}{c}{2nd-order correction} & 
  \multicolumn{1}{c}{3rd-order correction} \\
  \hline
  
  \ce{H4C} & -14.04 & -9.45 & -13.01 & -13.84 & -14.08 \\
  \ce{H3N} & -11.31 & -6.34 & -9.59 & -10.73 & -11.11 \\
  \ce{HO} & -13.50 & -7.41 & -11.10 & -12.45 & -13.04 \\
  \ce{H2O} & -13.15 & -7.38 & -11.07 & -12.38 & -12.87 \\
  \ce{HF} & -16.72 & -9.79 & -14.27 & -15.83 & -16.43 \\
  \ce{H4Si} & -12.15 & -8.51 & -11.59 & -12.32 & -12.46 \\
  \ce{HP} & -10.28 & -6.16 & -9.30 & -9.91 & -10.21 \\
  \ce{H2P} & -9.98 & -5.99 & -9.00 & -9.61 & -9.88 \\
  \ce{H3P} & -10.64 & -6.77 & -9.65 & -10.34 & -10.49 \\
  \ce{HS} & -10.58 & -6.29 & -9.06 & -9.83 & -10.13 \\
  \ce{H2S} (\ce{^2B_1}) & -10.61 & -6.39 & -9.32 & -10.09 & -10.35 \\
  \ce{H2S} (\ce{^2A_1}) & -13.39 & -9.09 & -12.32 & -13.08 & -13.27 \\
  \ce{HCl} & -12.97 & -8.15 & -11.55 & -12.42 & -12.71 \\
  \ce{H2C2} & -11.74 & -7.37 & -10.50 & -11.29 & -11.54 \\
  \ce{H4C2} & -11.00 & -6.96 & -9.92 & -10.64 & -10.84 \\
  \ce{CO} & -14.11 & -9.15 & -12.80 & -13.95 & -14.24 \\
  \ce{N2} (\ce{^2\Sigma_g}) & -15.63 & -10.43 & -14.30 & -15.23 & -15.52 \\
  \ce{N2} (\ce{^2\Pi_u}) & -17.39 & -11.85 & -15.98 & -16.96 & -17.24 \\
  \ce{O2} & -12.86 & -7.04 & -11.46 & -12.48 & -12.82 \\
  \ce{P2} & -10.70 & -7.24 & -9.88 & -10.42 & -10.56 \\
  \ce{S2} & -9.62 & -5.94 & -8.93 & -9.43 & -9.57 \\
  \ce{Cl2} & -11.47 & -7.56 & -10.71 & -11.27 & -11.40 \\
  \ce{FCl} & -12.75 & -8.11 & -11.67 & -12.41 & -12.59 \\
  \ce{CS} & -11.51 & -7.49 & -10.54 & -11.61 & -11.91 \\
  \ce{BF3} & -14.80 & -10.34 & -14.03 & -14.74 & -14.84 \\
  \ce{BCl3} & -10.99 & -7.87 & -10.54 & -10.93 & -10.99 \\
  \ce{B2F4} & -12.43 & -8.64 & -11.85 & -12.47 & -12.43 \\
  \ce{CO2} & -14.03 & -9.33 & -13.10 & -13.95 & -14.13 \\
  \ce{CF2} & -12.15 & -7.49 & -10.90 & -11.87 & -12.12 \\
  \ce{COS} & -11.49 & -7.65 & -10.72 & -11.35 & -11.46 \\
  \ce{CS2} & -10.24 & -6.95 & -9.66 & -10.15 & -10.25 \\
  \ce{H2C} & -10.71 & -5.76 & -9.27 & -10.11 & -10.66 \\
  \ce{H3C} & -10.09 & -5.40 & -8.71 & -9.51 & -10.02 \\
  \ce{H5C2} & -8.54 & -4.55 & -7.57 & -8.34 & -8.60 \\
  \ce{H4C3} (cyc) & -9.90 & -6.23 & -9.14 & -9.68 & -9.80 \\
  \ce{H4C3} (all) & -10.25 & -6.72 & -9.53 & -10.19 & -10.27 \\
  \ce{H7C3} & -7.60 & -4.02 & -6.80 & -7.56 & -7.74 \\
  \ce{H6C6} & -9.56 & -6.54 & -9.03 & -9.53 & -9.59 \\
  \ce{H8C7} & -9.01 & -6.19 & -8.54 & -9.00 & -9.02 \\
  \ce{CN} & -14.57 & -9.54 & -13.00 & -13.92 & -14.20 \\
  \ce{HCO} & -9.57 & -4.97 & -8.41 & -9.10 & -9.47 \\
  \ce{CH2OH} & -8.27 & -3.97 & -7.14 & -7.89 & -8.28 \\
  \ce{CH3O} & -10.64 & -6.04 & -9.34 & -10.39 & -10.49 \\
  \ce{H4CO} & -10.89 & -6.41 & -9.74 & -10.75 & -10.71 \\
  \ce{H3CF} & -12.82 & -8.18 & -11.80 & -12.69 & -12.54 \\
  \ce{H2CS} & -9.45 & -5.63 & -8.46 & -9.17 & -9.34 \\
  \ce{CH2SH} & -7.83 & -4.19 & -7.03 & -7.63 & -7.86 \\
  \ce{H4CS} & -9.50 & -5.70 & -8.49 & -9.22 & -9.36 \\
  \ce{H3CCl} & -11.30 & -7.20 & -10.32 & -11.11 & -11.14 \\
  \ce{H6C2O} & -10.30 & -6.33 & -9.41 & -10.39 & -10.18 \\
  \ce{H4C2O} & -10.24 & -6.09 & -9.13 & -10.13 & -10.30 \\
  \ce{H3COF} & -11.16 & -6.77 & -10.17 & -11.15 & -11.14 \\
  \ce{H4C2S} & -9.15 & -5.48 & -8.21 & -8.94 & -9.08 \\
  \ce{C2N2} & -13.37 & -9.56 & -12.70 & -13.28 & -13.37 \\
  \ce{H4C4O} & -9.23 & -5.90 & -8.59 & -9.15 & -9.24 \\
  \ce{H5C4N} & -8.60 & -5.37 & -7.96 & -8.52 & -8.62 \\
  \ce{H6C6O} & -8.67 & -5.76 & -8.19 & -8.64 & -8.68 \\
  \ce{H7C6N} & -8.04 & -5.21 & -7.56 & -8.02 & -8.05 \\
  \ce{H4B2} & -10.32 & -6.47 & -9.29 & -10.02 & -10.20 \\
  \ce{HN} & -13.98 & -7.98 & -12.15 & -13.26 & -13.87 \\
  \ce{H2N} & -12.38 & -7.27 & -10.55 & -11.72 & -12.07 \\
  \ce{H2N2} & -10.17 & -5.72 & -9.00 & -9.73 & -9.99 \\
  \ce{H3N2} & -8.57 & -4.17 & -7.41 & -8.21 & -8.52 \\
  \ce{HOF} & -12.90 & -7.49 & -11.51 & -12.56 & -12.83 \\
  \ce{H2Si} & -9.49 & -5.91 & -8.47 & -9.06 & -9.22 \\
  \ce{H3Si} & -8.92 & -5.30 & -8.11 & -8.63 & -8.87 \\
  \ce{H2Si2} & -8.19 & -5.21 & -7.42 & -7.89 & -8.02 \\
  \ce{H4Si2} & -8.42 & -5.53 & -7.73 & -8.17 & -8.27 \\
  \ce{H5Si2} & -8.34 & -5.21 & -7.71 & -8.25 & -8.39 \\
  \ce{H6Si2} & -10.49 & -7.36 & -9.91 & -10.45 & -10.49 \\
  \hline\hline
  \multicolumn{1}{c}{MAD} & - & 4.19 & 1.05 & 0.29 & 0.12 \\
  \hline
\end{longtable}

\begin{longtable}[c]{cccccc}
  \caption{MAD of EAs for molecules in G2-97 datasets with uncorrected LDA and modified formalism for perturbative LDA exchange potential to different orders.}\label{g2_lda_ea}\\
  \hline

  \multicolumn{1}{c}{molecules} & 
  \multicolumn{1}{c}{-A} & 
  \multicolumn{1}{c}{Original method} & 
  \multicolumn{1}{c}{1st-order correction} & 
  \multicolumn{1}{c}{2nd-order correction} & 
  \multicolumn{1}{c}{3rd-order correction} \\
  \hline
  
  \ce{HC} & -1.70 & -5.79 & -2.02 & -1.14 & -1.73 \\
  \ce{H2C} & -0.55 & -3.90 & -0.90 & -0.02 & -0.53 \\
  \ce{H3C} & -0.29 & -3.59 & -0.47 & 0.33 & -0.20 \\
  \ce{HN} & -0.71 & -4.87 & -1.10 & -0.05 & -0.69 \\
  \ce{H2N} & -1.12 & -5.23 & -1.49 & -0.47 & -1.01 \\
  \ce{HO} & -2.26 & -7.25 & -2.81 & -1.54 & -2.15 \\
  \ce{HSi} & -1.44 & -4.44 & -1.57 & -1.09 & -1.39 \\
  \ce{H2Si} & -1.29 & -4.25 & -1.44 & -0.97 & -1.26 \\
  \ce{H3Si} & -1.15 & -3.98 & -1.42 & -0.87 & -1.09 \\
  \ce{HP} & -1.25 & -4.42 & -1.61 & -0.97 & -1.28 \\
  \ce{H2P} & -1.47 & -4.67 & -1.76 & -1.14 & -1.41 \\
  \ce{HS} & -2.57 & -6.28 & -2.97 & -2.21 & -2.51 \\
  \ce{O2} & -0.09 & -4.97 & -0.58 & 0.41 & 0.05 \\
  \ce{NO} & -0.01 & -4.65 & -0.39 & 0.53 & 0.11 \\
  \ce{CN} & -4.01 & -8.12 & -4.40 & -3.51 & -3.87 \\
  \ce{OP} & -1.22 & -4.65 & -1.39 & -0.86 & -1.12 \\
  \ce{S2} & -1.62 & -4.88 & -1.92 & -1.43 & -1.58 \\
  \ce{Cl2} & -1.37 & -4.92 & -1.68 & -1.10 & -1.30 \\
  \ce{C2} & -4.16 & -8.10 & -4.46 & -3.66 & -3.95 \\
  \ce{C2O} & -2.37 & -6.10 & -2.77 & -2.13 & -2.28 \\
  \ce{CF2} & 0.03 & -4.07 & -0.31 & 0.58 & 0.25 \\
  \ce{CNO} & -3.72 & -7.81 & -4.17 & -3.41 & -3.57 \\
  \ce{NO2} & -1.49 & -5.57 & -1.85 & -1.08 & -1.34 \\
  \ce{O3} & -1.96 & -6.41 & -2.50 & -1.71 & -1.89 \\
  \ce{OF} & -2.07 & -7.19 & -2.69 & -1.57 & -1.86 \\
  \ce{O2S} & -1.10 & -4.89 & -1.47 & -0.89 & -1.05 \\
  \ce{OS2} & -1.85 & -5.15 & -2.17 & -1.71 & -1.82 \\
  \ce{HC2} & -3.20 & -6.89 & -3.68 & -2.60 & -2.86 \\
  \ce{H3C2} & -0.43 & -3.68 & -0.72 & 0.08 & -0.21 \\
  \ce{H2C3} & -2.00 & -5.35 & -2.34 & -1.76 & -1.93 \\
  \ce{H3C3} & -1.00 & -4.10 & -1.27 & -0.66 & -0.84 \\
  \ce{H5C3} & -0.73 & -3.72 & -1.00 & -0.41 & -0.58 \\
  \ce{HCO} & -0.12 & -3.65 & -0.23 & 0.48 & 0.07 \\
  \ce{HCF} & -0.59 & -4.55 & -0.90 & -0.05 & -0.50 \\
  \ce{CH3O} & -1.67 & -5.85 & -2.27 & -1.28 & -1.37 \\
  \ce{H3CS} & 0.00 & -3.06 & -0.25 & 0.37 & 0.15 \\
  \ce{H2CS} & -0.52 & -3.84 & -0.79 & -0.25 & -0.47 \\
  \ce{CH2CN} & -1.85 & -5.24 & -2.18 & -1.51 & -1.68 \\
  \ce{CH2NC} & -1.41 & -4.68 & -1.72 & -1.07 & -1.26 \\
  \ce{HC2O} & -2.47 & -6.13 & -2.86 & -2.20 & -2.35 \\
  \ce{CH2CHO} & -2.06 & -5.48 & -2.43 & -1.76 & -1.89 \\
  \ce{CH3CO} & -0.44 & -3.27 & -0.38 & 0.38 & -0.15 \\
  \ce{H5C2O} & -1.95 & -5.84 & -2.69 & -1.68 & -1.65 \\
  \ce{H5C2S} & -2.09 & -5.40 & -2.57 & -1.82 & -1.93 \\
  \ce{HLi} & -0.41 & -1.55 & -0.14 & -0.01 & -0.54 \\
  \ce{HNO} & -0.32 & -4.72 & -0.78 & 0.10 & -0.25 \\
  \ce{HO2} & -0.68 & -5.31 & -1.23 & -0.23 & -0.51 \\
  \hline\hline
  \multicolumn{1}{c}{MAD} & - & 3.65 & 0.35 & 0.41 & 0.13 \\
  \hline
\end{longtable}

\begin{longtable}[c]{cccccc}
  \caption{MAD of IPs for molecules in G2-97 datasets with uncorrected BLYP and modified formalism for perturbative BLYP exchange potential to different orders.}\label{g2_blyp_ip}\\
  \hline

  \multicolumn{1}{c}{molecules} & 
  \multicolumn{1}{c}{-I} & 
  \multicolumn{1}{c}{Original method} & 
  \multicolumn{1}{c}{1st-order correction} & 
  \multicolumn{1}{c}{2nd-order correction} & 
  \multicolumn{1}{c}{3rd-order correction} \\
  \hline
  
  \ce{H4C} & -13.87 & -9.37 & -12.92 & -13.78 & -14.00 \\
  \ce{H3N} & -10.98 & -6.17 & -9.35 & -10.52 & -10.91 \\
  \ce{HO} & -13.29 & -7.36 & -10.97 & -12.36 & -12.97 \\
  \ce{H2O} & -12.73 & -7.19 & -10.80 & -12.13 & -12.64 \\
  \ce{HF} & -16.26 & -9.60 & -13.98 & -15.57 & -16.18 \\
  \ce{H4Si} & -12.05 & -8.43 & -11.52 & -12.28 & -12.35 \\
  \ce{HP} & -9.97 & -5.89 & -9.01 & -9.63 & -9.94 \\
  \ce{H2P} & -9.73 & -5.78 & -8.79 & -9.40 & -9.68 \\
  \ce{H3P} & -10.42 & -6.60 & -9.45 & -10.16 & -10.32 \\
  \ce{HS} & -10.32 & -6.10 & -8.80 & -9.59 & -9.91 \\
  \ce{H2S} (\ce{^2B_1}) & -10.27 & -6.15 & -9.04 & -9.81 & -10.10 \\
  \ce{H2S} (\ce{^2A_1}) & -13.14 & -8.92 & -12.12 & -12.89 & -13.09 \\
  \ce{HCl} & -12.60 & -7.91 & -11.26 & -12.14 & -12.45 \\
  \ce{H2C2} & -11.28 & -7.03 & -10.10 & -10.89 & -11.17 \\
  \ce{H4C2} & -10.53 & -6.61 & -9.51 & -10.24 & -10.46 \\
  \ce{CO} & -13.91 & -9.04 & -12.65 & -13.81 & -14.12 \\
  \ce{N2} (\ce{^2\Sigma_g}) & -15.37 & -10.26 & -14.08 & -15.02 & -15.33 \\
  \ce{N2} (\ce{^2\Pi_u}) & -16.82 & -11.45 & -15.51 & -16.50 & -16.79 \\
  \ce{O2} & -12.59 & -6.89 & -11.24 & -12.28 & -12.64 \\
  \ce{P2} & -10.29 & -6.89 & -9.48 & -10.03 & -10.17 \\
  \ce{S2} & -9.35 & -5.72 & -8.68 & -9.18 & -9.33 \\
  \ce{Cl2} & -11.15 & -7.30 & -10.41 & -10.98 & -11.11 \\
  \ce{FCl} & -12.42 & -7.87 & -11.38 & -12.13 & -12.32 \\
  \ce{CS} & -11.27 & -7.34 & -10.34 & -11.42 & -11.74 \\
  \ce{BF3} & -14.40 & -10.04 & -13.69 & -14.41 & -14.51 \\
  \ce{BCl3} & -10.64 & -7.57 & -10.21 & -10.61 & -10.67 \\
  \ce{B2F4} & -12.12 & -8.40 & -11.58 & -12.21 & -12.16 \\
  \ce{CO2} & -13.54 & -8.99 & -12.70 & -13.57 & -13.75 \\
  \ce{CF2} & -11.92 & -7.36 & -10.72 & -11.70 & -11.97 \\
  \ce{COS} & -11.07 & -7.32 & -10.34 & -10.98 & -11.09 \\
  \ce{CS2} & -9.84 & -6.63 & -9.30 & -9.79 & -9.90 \\
  \ce{H2C} & -10.32 & -5.50 & -8.96 & -9.82 & -10.41 \\
  \ce{H3C} & -9.80 & -5.22 & -8.50 & -9.31 & -9.87 \\
  \ce{H5C2} & -8.36 & -4.43 & -7.45 & -8.24 & -8.52 \\
  \ce{H4C3} (cyc) & -9.58 & -5.98 & -8.84 & -9.41 & -9.54 \\
  \ce{H4C3} (all) & -9.88 & -6.42 & -9.19 & -9.87 & -9.97 \\
  \ce{H7C3} & -7.47 & -3.94 & -6.73 & -7.51 & -7.70 \\
  \ce{H6C6} & -9.10 & -6.14 & -8.61 & -9.11 & -9.18 \\
  \ce{H8C7} & -8.60 & -5.82 & -8.16 & -8.63 & -8.66 \\
  \ce{CN} & -14.15 & -9.26 & -12.67 & -13.60 & -13.90 \\
  \ce{HCO} & -9.49 & -4.94 & -8.36 & -9.06 & -9.44 \\
  \ce{CH2OH} & -8.08 & -3.85 & -6.99 & -7.75 & -8.17 \\
  \ce{CH3O} & -10.52 & -6.03 & -9.31 & -10.39 & -10.48 \\
  \ce{H4CO} & -10.65 & -6.26 & -9.56 & -10.61 & -10.57 \\
  \ce{H3CF} & -12.63 & -8.04 & -11.65 & -12.59 & -12.38 \\
  \ce{H2CS} & -9.15 & -5.41 & -8.20 & -8.92 & -9.11 \\
  \ce{CH2SH} & -7.61 & -4.03 & -6.84 & -7.44 & -7.68 \\
  \ce{H4CS} & -9.21 & -5.49 & -8.23 & -8.98 & -9.14 \\
  \ce{H3CCl} & -11.01 & -6.97 & -10.06 & -10.87 & -10.93 \\
  \ce{H6C2O} & -10.14 & -6.17 & -9.26 & -10.29 & -10.07 \\
  \ce{H4C2O} & -9.96 & -5.92 & -8.92 & -9.95 & -9.91 \\
  \ce{H3COF} & -10.94 & -6.62 & -10.01 & -11.01 & -11.02 \\
  \ce{H4C2S} & -8.81 & -5.23 & -7.92 & -8.68 & -8.83 \\
  \ce{C2N2} & -12.88 & -9.15 & -12.25 & -12.84 & -12.94 \\
  \ce{H4C4O} & -8.77 & -5.53 & -8.18 & -8.76 & -8.85 \\
  \ce{H5C4N} & -8.15 & -5.01 & -7.56 & -8.13 & -8.23 \\
  \ce{H6C6O} & -8.25 & -5.39 & -7.80 & -8.25 & -8.31 \\
  \ce{H7C6N} & -7.64 & -4.87 & -7.20 & -7.65 & -7.69 \\
  \ce{H4B2} & -9.99 & -6.22 & -9.00 & -9.75 & -9.94 \\
  \ce{HN} & -13.57 & -7.73 & -11.86 & -12.99 & -13.63 \\
  \ce{H2N} & -12.26 & -7.23 & -10.46 & -11.66 & -12.03 \\
  \ce{H2N2} & -10.01 & -5.64 & -8.87 & -9.62 & -9.88 \\
  \ce{H3N2} & -8.39 & -4.08 & -7.27 & -8.08 & -8.41 \\
  \ce{HOF} & -12.59 & -7.32 & -11.28 & -12.34 & -12.62 \\
  \ce{H2Si} & -9.31 & -5.76 & -8.30 & -8.90 & -9.07 \\
  \ce{H3Si} & -8.74 & -5.13 & -7.94 & -8.47 & -8.73 \\
  \ce{H2Si2} & -7.85 & -4.92 & -7.09 & -7.56 & -7.70 \\
  \ce{H4Si2} & -8.09 & -5.24 & -7.41 & -7.86 & -7.96 \\
  \ce{H5Si2} & -8.14 & -5.01 & -7.52 & -8.07 & -8.22 \\
  \ce{H6Si2} & -10.26 & -7.16 & -9.72 & -10.28 & -10.32 \\
  \hline\hline
  \multicolumn{1}{c}{MAD} & - & 4.10 & 1.01 & 0.24 & 0.10 \\
  \hline
\end{longtable}

\begin{longtable}[c]{cccccc}
  \caption{MAD of EAs for molecules in G2-97 datasets with uncorrected BLYP and modified formalism for perturbative BLYP exchange potential to different orders.}\label{g2_blyp_ea}\\
  \hline

  \multicolumn{1}{c}{molecules} & 
  \multicolumn{1}{c}{-A} &  
  \multicolumn{1}{c}{Original method} & 
  \multicolumn{1}{c}{1st-order correction} & 
  \multicolumn{1}{c}{2nd-order correction} & 
  \multicolumn{1}{c}{3rd-order correction} \\
  \hline
  
  \ce{HC} & -1.29 & -5.16 & -1.47 & -0.62 & -1.27 \\
  \ce{H2C} & -0.37 & -3.59 & -0.67 & 0.19 & -0.37 \\
  \ce{H3C} & 0.06 & -3.06 & -0.04 & 0.70 & 0.10 \\
  \ce{HN} & -0.49 & -4.50 & -0.81 & 0.21 & -0.48 \\
  \ce{H2N} & -0.72 & -4.61 & -0.97 & 0.01 & -0.58 \\
  \ce{HO} & -1.83 & -6.53 & -2.19 & -0.96 & -1.62 \\
  \ce{HSi} & -1.11 & -4.03 & -1.21 & -0.73 & -1.07 \\
  \ce{H2Si} & -1.01 & -3.91 & -1.14 & -0.69 & -1.00 \\
  \ce{H3Si} & -0.86 & -3.59 & -1.08 & -0.54 & -0.79 \\
  \ce{HP} & -1.00 & -4.09 & -1.32 & -0.70 & -1.03 \\
  \ce{H2P} & -1.14 & -4.24 & -1.38 & -0.77 & -1.07 \\
  \ce{HS} & -2.23 & -5.81 & -2.54 & -1.80 & -2.12 \\
  \ce{O2} & 0.13 & -4.60 & -0.26 & 0.72 & 0.33 \\
  \ce{NO} & 0.26 & -4.25 & -0.05 & 0.87 & 0.41 \\
  \ce{CN} & -3.76 & -7.76 & -4.09 & -3.21 & -3.59 \\
  \ce{OP} & -0.96 & -4.33 & -1.11 & -0.57 & -0.85 \\
  \ce{S2} & -1.34 & -4.53 & -1.61 & -1.11 & -1.28 \\
  \ce{Cl2} & -1.12 & -4.59 & -1.40 & -0.83 & -1.05 \\
  \ce{C2} & -3.45 & -7.81 & -4.21 & -3.42 & -3.74 \\
  \ce{C2O} & -2.03 & -5.66 & -2.37 & -1.73 & -1.90 \\
  \ce{CF2} & 0.24 & -3.68 & 0.02 & 0.91 & 0.53 \\
  \ce{CNO} & -3.26 & -7.20 & -3.60 & -2.86 & -3.03 \\
  \ce{NO2} & -1.23 & -5.19 & -1.53 & -0.76 & -1.05 \\
  \ce{O3} & -1.74 & -6.09 & -2.22 & -1.42 & -1.62 \\
  \ce{OF} & -1.71 & -6.61 & -2.18 & -1.06 & -1.38 \\
  \ce{O2S} & -0.88 & -4.58 & -1.20 & -0.62 & -0.79 \\
  \ce{OS2} & -1.62 & -4.84 & -1.90 & -1.44 & -1.55 \\
  \ce{HC2} & -2.93 & -6.49 & -3.26 & -2.31 & -2.56 \\
  \ce{H3C2} & -0.18 & -3.26 & -0.36 & 0.43 & 0.09 \\
  \ce{H2C3} & -1.66 & -4.89 & -1.92 & -1.35 & -1.54 \\
  \ce{H3C3} & -0.61 & -3.57 & -0.79 & -0.20 & -0.40 \\
  \ce{H5C3} & -0.34 & -3.19 & -0.52 & 0.06 & -0.13 \\
  \ce{HCO} & 0.06 & -3.36 & 0.02 & 0.71 & 0.26 \\
  \ce{HCF} & -0.34 & -4.12 & -0.54 & 0.29 & -0.21 \\
  \ce{CH3O} & -1.33 & -5.33 & -1.77 & -0.75 & -0.88 \\
  \ce{H3CS} & 0.32 & -2.60 & 0.15 & 0.76 & 0.51 \\
  \ce{H2CS} & -0.28 & -3.50 & -0.50 & 0.05 & -0.21 \\
  \ce{CH2CN} & -1.44 & -4.67 & -1.67 & -1.01 & -1.20 \\
  \ce{CH2NC} & -1.00 & -4.11 & -1.20 & -0.57 & -0.79 \\
  \ce{HC2O} & -2.06 & -5.58 & -2.36 & -1.71 & -1.88 \\
  \ce{CH2CHO} & -1.65 & -4.92 & -1.92 & -1.26 & -1.42 \\
  \ce{CH3CO} & -0.22 & -2.95 & -0.13 & 0.59 & -0.09 \\
  \ce{H5C2O} & -1.58 & -5.20 & -2.03 & -0.99 & -0.99 \\
  \ce{H5C2S} & -1.76 & -4.96 & -2.16 & -1.40 & -1.59 \\
  \ce{HLi} & -0.34 & -1.71 & -0.23 & -0.09 & -0.69 \\
  \ce{HNO} & -0.10 & -4.35 & -0.47 & 0.41 & 0.03 \\
  \ce{HO2} & -0.37 & -4.82 & -0.80 & 0.20 & -0.10 \\
  \hline\hline
  \multicolumn{1}{c}{MAD} & - & 3.54 & 0.28 & 0.47 & 0.17 \\
  \hline
\end{longtable}

\begin{longtable}[c]{cccccc}
  \caption{MAD of IPs for molecules in G2-97 datasets with uncorrected B3LYP and modified formalism for perturbative B3LYP exchange potential to different orders.}\label{g2_b3lyp_ip}\\
  \hline

  \multicolumn{1}{c}{molecules} & 
  \multicolumn{1}{c}{-I} & 
  \multicolumn{1}{c}{Original method} & 
  \multicolumn{1}{c}{1st-order correction} & 
  \multicolumn{1}{c}{2nd-order correction} & 
  \multicolumn{1}{c}{3rd-order correction} \\
  \hline
  
  \ce{H4C} & -14.14 & -10.75 & -13.11 & -14.07 & -14.20 \\
  \ce{H3N} & -11.06 & -7.57 & -9.38 & -10.83 & -10.99 \\
  \ce{HO} & -13.32 & -9.00 & -11.05 & -12.69 & -13.04 \\
  \ce{H2O} & -12.77 & -8.81 & -10.80 & -12.44 & -12.69 \\
  \ce{HF} & -16.27 & -11.52 & -13.97 & -15.85 & -16.20 \\
  \ce{H4Si} & -12.45 & -9.66 & -11.73 & -12.58 & -12.64 \\
  \ce{HP} & -10.16 & -7.02 & -9.33 & -9.97 & -10.15 \\
  \ce{H2P} & -9.92 & -6.89 & -9.09 & -9.72 & -9.89 \\
  \ce{H3P} & -10.58 & -7.68 & -9.61 & -10.42 & -10.50 \\
  \ce{HS} & -10.46 & -7.25 & -9.08 & -9.96 & -10.15 \\
  \ce{H2S} (\ce{^2B_1}) & -10.42 & -7.32 & -9.26 & -10.14 & -10.30 \\
  \ce{H2S} (\ce{^2A_1}) & -13.34 & -10.15 & -12.29 & -13.17 & -13.30 \\
  \ce{HCl} & -12.76 & -9.23 & -11.47 & -12.47 & -12.65 \\
  \ce{H2C2} & -11.35 & -8.19 & -10.19 & -11.08 & -11.25 \\
  \ce{H4C2} & -10.57 & -7.66 & -9.56 & -10.38 & -10.50 \\
  \ce{CO} & -14.22 & -10.56 & -12.83 & -14.22 & -14.42 \\
  \ce{N2} (\ce{^2\Sigma_g}) & -15.83 & -11.97 & -14.56 & -15.62 & -15.82 \\
  \ce{N2} (\ce{^2\Pi_u}) & -16.90 & -12.85 & -15.58 & -16.66 & -16.86 \\
  \ce{O2} & -13.11 & -8.81 & -11.82 & -12.93 & -13.15 \\
  \ce{P2} & -10.41 & -7.83 & -9.62 & -10.23 & -10.32 \\
  \ce{S2} & -9.71 & -6.92 & -9.08 & -9.61 & -9.71 \\
  \ce{Cl2} & -11.55 & -8.61 & -10.82 & -11.45 & -11.54 \\
  \ce{FCl} & -12.82 & -9.35 & -11.80 & -12.64 & -12.76 \\
  \ce{CS} & -11.52 & -8.72 & -10.31 & -11.73 & -12.02 \\
  \ce{BF3} & -15.24 & -11.98 & -14.45 & -15.28 & -15.35 \\
  \ce{BCl3} & -11.24 & -8.88 & -10.78 & -11.23 & -11.27 \\
  \ce{B2F4} & -12.81 & -9.93 & -12.15 & -12.88 & -12.87 \\
  \ce{CO2} & -13.80 & -10.47 & -12.79 & -13.92 & -14.04 \\
  \ce{CF2} & -12.28 & -8.83 & -11.03 & -12.18 & -12.34 \\
  \ce{COS} & -11.29 & -8.46 & -10.47 & -11.27 & -11.33 \\
  \ce{CS2} & -10.05 & -7.63 & -9.40 & -10.08 & -10.15 \\
  \ce{H2C} & -10.45 & -6.81 & -9.30 & -10.19 & -10.51 \\
  \ce{H3C} & -9.91 & -6.47 & -8.81 & -9.67 & -9.93 \\
  \ce{H5C2} & -8.58 & -5.66 & -7.70 & -8.55 & -8.67 \\
  \ce{H4C3} (cyc) & -9.75 & -7.04 & -9.00 & -9.64 & -9.73 \\
  \ce{H4C3} (all) & -10.11 & -7.51 & -9.24 & -10.06 & -10.11 \\
  \ce{H7C3} & -7.73 & -5.14 & -6.94 & -7.79 & -7.86 \\
  \ce{H6C6} & -9.26 & -7.08 & -8.67 & -9.30 & -9.33 \\
  \ce{H8C7} & -8.80 & -6.75 & -8.25 & -8.84 & -8.86 \\
  \ce{CN} & -14.17 & -10.54 & -12.69 & -13.80 & -13.94 \\
  \ce{HCO} & -9.82 & -6.32 & -8.87 & -9.59 & -9.80 \\
  \ce{CH2OH} & -8.36 & -5.18 & -7.41 & -8.22 & -8.41 \\
  \ce{CH3O} & -10.82 & -7.60 & -9.48 & -10.76 & -10.85 \\
  \ce{H4CO} & -10.93 & -7.77 & -9.66 & -10.90 & -10.97 \\
  \ce{H3CF} & -13.05 & -9.67 & -11.98 & -12.97 & -13.05 \\
  \ce{H2CS} & -9.33 & -6.54 & -8.31 & -9.23 & -9.34 \\
  \ce{CH2SH} & -7.89 & -5.21 & -7.18 & -7.81 & -7.94 \\
  \ce{H4CS} & -9.41 & -6.63 & -8.42 & -9.28 & -9.38 \\
  \ce{H3CCl} & -11.29 & -8.26 & -10.27 & -11.19 & -11.26 \\
  \ce{H6C2O} & -10.54 & -7.65 & -9.39 & -10.58 & -10.61 \\
  \ce{H4C2O} & -10.20 & -7.35 & -8.96 & -10.24 & -10.10 \\
  \ce{H3COF} & -11.43 & -8.33 & -10.32 & -11.50 & -11.55 \\
  \ce{H4C2S} & -9.00 & -6.37 & -8.05 & -8.94 & -9.03 \\
  \ce{C2N2} & -13.14 & -10.34 & -12.39 & -13.17 & -13.22 \\
  \ce{H4C4O} & -8.91 & -6.51 & -8.22 & -8.93 & -8.98 \\
  \ce{H5C4N} & -8.28 & -5.96 & -7.59 & -8.30 & -8.35 \\
  \ce{H6C6O} & -8.50 & -6.38 & -7.96 & -8.53 & -8.56 \\
  \ce{H7C6N} & -7.88 & -5.83 & -7.34 & -7.92 & -7.95 \\
  \ce{H4B2} & -10.12 & -7.28 & -9.12 & -9.95 & -10.07 \\
  \ce{HN} & -13.70 & -9.37 & -12.13 & -13.35 & -13.72 \\
  \ce{H2N} & -12.31 & -8.62 & -10.44 & -11.88 & -12.07 \\
  \ce{H2N2} & -10.32 & -7.03 & -9.24 & -10.09 & -10.25 \\
  \ce{H3N2} & -8.71 & -5.52 & -7.66 & -8.55 & -8.72 \\
  \ce{HOF} & -13.02 & -9.15 & -11.66 & -12.91 & -13.12 \\
  \ce{H2Si} & -9.48 & -6.73 & -8.52 & -9.20 & -9.30 \\
  \ce{H3Si} & -8.92 & -6.14 & -8.24 & -8.77 & -8.90 \\
  \ce{H2Si2} & -8.04 & -5.80 & -7.31 & -7.87 & -7.96 \\
  \ce{H4Si2} & -8.18 & -6.01 & -7.51 & -8.03 & -8.08 \\
  \ce{H5Si2} & -8.36 & -5.99 & -7.75 & -8.33 & -8.40 \\
  \ce{H6Si2} & -10.54 & -8.21 & -9.90 & -10.56 & -10.59 \\
  \hline\hline
  \multicolumn{1}{c}{MAD} & - & 3.06 & 1.03 & 0.15 & 0.08 \\
  \hline
\end{longtable}

\begin{longtable}[c]{cccccc}
  \caption{MAD of EAs for molecules in G2-97 datasets with uncorrected B3LYP and modified formalism for perturbative B3LYP exchange potential to different orders.}\label{g2_b3lyp_ea}\\
  \hline

  \multicolumn{1}{c}{molecules} & 
  \multicolumn{1}{c}{-A} & 
  \multicolumn{1}{c}{Original method} & 
  \multicolumn{1}{c}{1st-order correction} & 
  \multicolumn{1}{c}{2nd-order correction} & 
  \multicolumn{1}{c}{3rd-order correction} \\
  \hline
  
  \ce{HC} & -1.34 & -4.15 & -1.81 & -0.93 & -1.33 \\
  \ce{H2C} & -0.34 & -2.69 & -0.95 & -0.12 & -0.52 \\
  \ce{H3C} & 0.07 & -2.20 & -0.34 & 0.36 & -0.01 \\
  \ce{HN} & -0.43 & -3.31 & -1.09 & -0.09 & -0.53 \\
  \ce{H2N} & -0.69 & -3.45 & -1.27 & -0.30 & -0.66 \\
  \ce{HO} & -1.75 & -5.03 & -2.49 & -1.24 & -1.66 \\
  \ce{HSi} & -1.24 & -3.44 & -1.53 & -1.01 & -1.23 \\
  \ce{H2Si} & -1.14 & -3.34 & -1.45 & -0.97 & -1.16 \\
  \ce{H3Si} & -0.95 & -3.01 & -1.38 & -0.83 & -1.03 \\
  \ce{HP} & -1.08 & -3.40 & -1.61 & -0.97 & -1.20 \\
  \ce{H2P} & -1.23 & -3.55 & -1.69 & -1.06 & -1.26 \\
  \ce{HS} & -2.32 & -4.97 & -2.86 & -2.08 & -2.31 \\
  \ce{O2} & -0.06 & -3.54 & -0.70 & 0.33 & 0.06 \\
  \ce{NO} & 0.06 & -3.28 & -0.50 & 0.46 & 0.14 \\
  \ce{CN} & -4.06 & -7.00 & -4.70 & -3.73 & -3.98 \\
  \ce{OP} & -1.18 & -3.73 & -1.51 & -0.93 & -1.11 \\
  \ce{S2} & -1.57 & -3.98 & -1.96 & -1.43 & -1.54 \\
  \ce{Cl2} & -1.30 & -3.90 & -1.75 & -1.13 & -1.28 \\
  \ce{C2} & -4.37 & -7.31 & -4.89 & -4.09 & -4.30 \\
  \ce{C2O} & -2.16 & -4.87 & -2.67 & -1.98 & -2.11 \\
  \ce{CF2} & 0.04 & -2.86 & -0.50 & 0.48 & 0.21 \\
  \ce{CNO} & -3.43 & -6.32 & -3.99 & -3.14 & -3.27 \\
  \ce{NO2} & -1.51 & -4.47 & -2.04 & -1.21 & -1.41 \\
  \ce{O3} & -2.29 & -5.52 & -2.93 & -2.04 & -2.19 \\
  \ce{OF} & -1.90 & -5.42 & -2.67 & -1.44 & -1.69 \\
  \ce{O2S} & -1.24 & -4.04 & -1.70 & -1.06 & -1.17 \\
  \ce{OS2} & -2.00 & -4.43 & -2.48 & -1.84 & -2.03 \\
  \ce{HC2} & -3.09 & -5.61 & -3.78 & -2.74 & -2.99 \\
  \ce{H3C2} & -0.22 & -2.43 & -0.75 & 0.09 & -0.14 \\
  \ce{H2C3} & -1.85 & -4.24 & -2.32 & -1.62 & -1.75 \\
  \ce{H3C3} & -0.70 & -2.86 & -1.12 & -0.43 & -0.56 \\
  \ce{H5C3} & -0.39 & -2.46 & -0.81 & -0.13 & -0.24 \\
  \ce{HCO} & -0.01 & -2.58 & -0.34 & 0.35 & 0.06 \\
  \ce{HCF} & -0.47 & -3.24 & -0.97 & -0.08 & -0.40 \\
  \ce{CH3O} & -1.34 & -4.12 & -2.07 & -0.92 & -1.08 \\
  \ce{H3CS} & -0.24 & -1.89 & -0.14 & 0.50 & 0.33 \\
  \ce{H2CS} & -0.48 & -2.88 & -0.91 & -0.26 & -0.46 \\
  \ce{CH2CN} & -1.55 & -3.92 & -2.03 & -1.27 & -1.40 \\
  \ce{CH2NC} & -1.03 & -3.31 & -1.46 & -0.76 & -0.91 \\
  \ce{HC2O} & -2.20 & -4.81 & -2.70 & -1.98 & -2.11 \\
  \ce{CH2CHO} & -1.70 & -4.08 & -2.21 & -1.45 & -1.57 \\
  \ce{CH3CO} & -0.20 & -2.20 & -0.46 & 0.25 & -0.23 \\
  \ce{H5C2O} & -1.54 & -4.05 & -2.30 & -1.11 & -1.21 \\
  \ce{H5C2S} & -1.86 & -4.20 & -2.49 & -1.65 & -1.82 \\
  \ce{HLi} & -0.43 & -1.41 & -0.41 & -0.28 & -0.66 \\
  \ce{HNO} & -0.27 & -3.37 & -0.89 & 0.06 & -0.19 \\
  \ce{HO2} & -0.52 & -3.72 & -1.22 & -0.12 & -0.34 \\
  \hline\hline
  \multicolumn{1}{c}{MAD} & - & 2.58 & 0.50 & 0.30 & 0.11 \\
  \hline
\end{longtable}

\end{document}